\documentclass{llncs}
\pdfoutput=1

\usepackage{amsmath,amssymb,verbatim}
\usepackage{float}

\newtheorem{open}{Open}
\usepackage{xspace}
\usepackage[ruled,vlined]{algorithm2e}
\usepackage{multirow}
\usepackage{mdwlist}
\usepackage{subfig}
\usepackage{graphicx}
\usepackage[font=small,font=footnotesize,textfont=it,labelfont=bf]{caption}
\usepackage{siunitx}
\usepackage{color}
\usepackage{graphicx,epstopdf}
\usepackage[numbers,square]{natbib}
\usepackage{url}

\newcommand{\eps}{\varepsilon}               

\newcommand{\marrow}{\marginpar[\hfill$\longrightarrow$]{$\longleftarrow$}}
\newcommand{\todo}[1]{\textcolor{blue}{\textbf{TODO:} \marrow{\textsf{#1}}}}
\newcommand{\etal}{\emph{et al.}\xspace}

\newcommand{\matr}[1]{\mathbf{#1}}
\renewcommand{\vec}[1]{\mathbf{#1}}

\newcommand{\citeN}[1]{\citet{#1}}
\renewcommand{\cite}[1]{\citep{#1}}
\newcommand{\shortcite}[1]{\shortcites{#1}}

\begin{document}

\title{Spatio-Temporal Analysis of Team Sports -- \\ A Survey}

\author{Joachim Gudmundsson 
\thanks{Joachim Gudmundsson was supported under the Australian Research Council's \emph{Discovery Projects} funding scheme (DP150101134).}
       \and
        Michael Horton 
    }
        
\institute{The University of Sydney, Australia}

\maketitle

\begin{abstract}

Team-based \emph{invasion sports} such as football, basketball and hockey are similar in the sense that the players are able to move freely around the playing area; and that player and team performance cannot be fully analysed without considering the movements and interactions of all players as a group.
State of the art object tracking systems now produce \emph{spatio-temporal} traces of player trajectories with high definition and high frequency, and this, in turn, has facilitated a variety of research efforts, across many disciplines, to extract insight from the trajectories.
We survey recent research efforts that use spatio-temporal data from team sports as input, and involve non-trivial computation.
This article categorises the research efforts in a coherent framework and identifies a number of open research questions.

\end{abstract}

\section{Introduction}
\label{sec:introduction}

Team sports are a significant recreational activity in many societies, and attract participants to compete in, watch, and also to capitalise from the sport. 
There are several sporting codes that can be classed together as \emph{invasion sports} in that they share a common structure: two teams are competing for possession of a ball (or puck) in a constrained playing area, for a given period of time, and each team has simultaneous objectives of scoring by putting the ball into the opposition's goal, and also defending their goal against attacks by the opposition.
The team that has scored the greatest number of goals at the end of the allotted time is the winner.
Football (soccer), basketball, ice hockey, field hockey, rugby, Australian Rules football, American football, handball, and Lacrosse are all examples of invasion sports.

Teams looking to improve their chances of winning will naturally seek to understand their performance, and also that of their opposition. 
Systematic analysis of sports play has been occurring since the $1950$s using manual notation methods~\cite{reep-1968}. 
However human observation can be unreliable -- experimental results in \citeN{franks-1986} 
      showed that the expert observers' recollection of significant match events 
is as low as \SI{42}{\percent} -- and in recent years, automated systems to capture and analyse sports play have proliferated.

Today, there are a number of systems in use that capture spatio-temporal data from team sports during matches.
The adoption of this technology and the availability to researchers of the resulting data varies amongst the different sporting codes and is driven by various factors, particularly commercial and technical.
There is a cost associated with installing and maintaining such systems, and while some leagues mandate that all stadium have systems fitted, in others the individual teams will bear the cost, and thus view the data as commercially sensitive.
Furthermore, the nature of some sports present technical challenges to automated systems, for example, sports such as rugby and American football have frequent collisions that can confound optical systems that rely on edge-detection.

\begin{figure}
\begin{center}
    \includegraphics[scale=1.0]{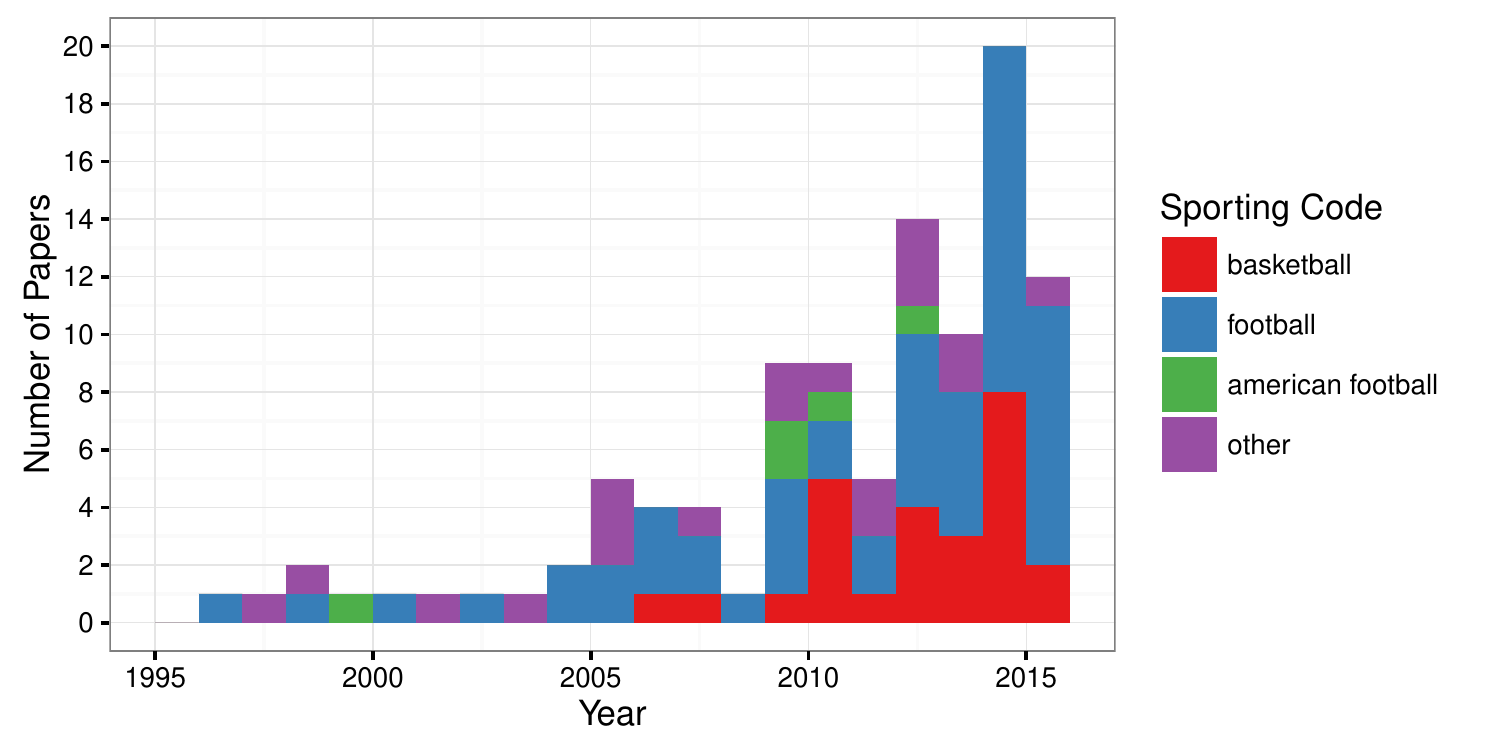}
\end{center}
\caption{Spatial sports research papers cited in this survey, by year, $1995$-$2015$ (to date), divided by sporting code. There has been a significant increase in papers published in this area as data has become available for researchers, particularly in football and basketball.}
\label{fig:references-histogram}
\end{figure}

To date, the majority of datasets available for research are sourced from football and basketball, and the research we surveyed reflects this, see Fig.~\ref{fig:references-histogram}.
The National Hockey League intends to install a player tracking system for the $2015\mbox{/}6$ season, and this may precipitate research in ice hockey in coming years~\cite{sportvision}.

Sports performance is actively researched in a variety of disciplines.
To be explicit, the research that we consider in this survey fulfils three key criteria:

\begin{enumerate}
    \item We consider \textbf{team-based invasion sports}.
    \item The model used in the research has \textbf{spatio-temporal data} as its primary input.
    \item The model performs some \textbf{non-trivial computation} on the spatio-temporal data. 
\end{enumerate}

The research covered has come from many research communities, including machine learning, network science, GIS, computational geometry, computer vision, complex systems science, statistics and sports science.
There has been a consequent diversity of methods and models used in the research, and our intention in writing this survey was to provide an overview and framework on the research efforts to date.

Furthermore, the spatio-temporal data extracted from sports has several useful properties that make it convenient for fundamental research. For instance, player trajectories exhibit small spatial and temporal range, dense sampling rates, a small number of agents (i.e. players), highly cooperative and adversarial interaction between agents, and a latent structure.
As such, we believe that this survey is a timely contribution to this emerging area of research.

This survey contains the following sections. In Section~\ref{sec:representation} we describe the primary types of spatio-temporal data captured from team sports. 
We describe the properties of these data and outline the sports from which it is currently available.

Section~\ref{sec:subdivision} describes approaches that have been used to subdivide the playing area into regions that have a particular property.
The playing area may be discretized into a fixed subdivision and the occurrences of some phenomena counted, for instance a player occupying a particular region or a shot at goal occurring from that region, producing an \emph{intensity map} of the playing area. 
On the other hand, a subdivision of the playing area based on areas that are dominated by particular players has also been used in several papers.

In Section~\ref{sec:networks}, we survey approaches that represent temporal sequences of events as \emph{networks} and apply network-theoretic measure to them. For example, sequences of passes between players can be represented as a network with players as the vertices and weighted edges for the frequency of passes between pairs of players, and network measures be computed to quantify the passing performance.

Section~\ref{sec:data-mining} provides a task-oriented survey of the approaches to uncover information inherent in the spatio-temporal data using \emph{data mining} techniques. 
Furthermore, several papers define metrics to 
      measure 
the performance of players and teams, and these are discussed in Section~\ref{sec:performance-metrics}.

Finally, in Section~\ref{sec:visualisation}, we detail the research into \emph{visualisation} techniques to succinctly present metrics of sports performance.

\section{Representing Sports Play using Spatio-Temporal Data}
\label{sec:representation}

The research surveyed in this paper is based on \emph{spatio-temporal data}, the defining characteristic of which is that it is a sequence of samples containing the time-stamp and location of some phenomena.
In the team sports domain, two types of spatio-temporal data are commonly captured: \emph{object trajectories} that capture the movement of players or the ball; and \emph{event logs} that record the location and time of match events, such as passes, shots at goal or fouls.
These datasets, described in detail below, individually facilitate the spatiotemporal analysis of play, however they are complementary in that they describe different aspects of play, and can provide a richer explanation of the game when used in combination.
For example, the spatial formation in which a team arranges itself in will be apparent in the set of player trajectories.
However, the particular formation used may depend on whether the team is in possession of the ball, which can be determined from the event log.
On the other hand, a \emph{shot at goal} event contains the location from where the shot was made, but this may not be sufficient to make a qualitative rating of the shot. 
Such a rating ought to consider whether the shooter was closely marked by the defence, and the proximity of defenders -- properties that can be interpolated from the player trajectories.

\subsection{Object Trajectories} 

\begin{figure}
\begin{center}
    \includegraphics[scale=0.7]{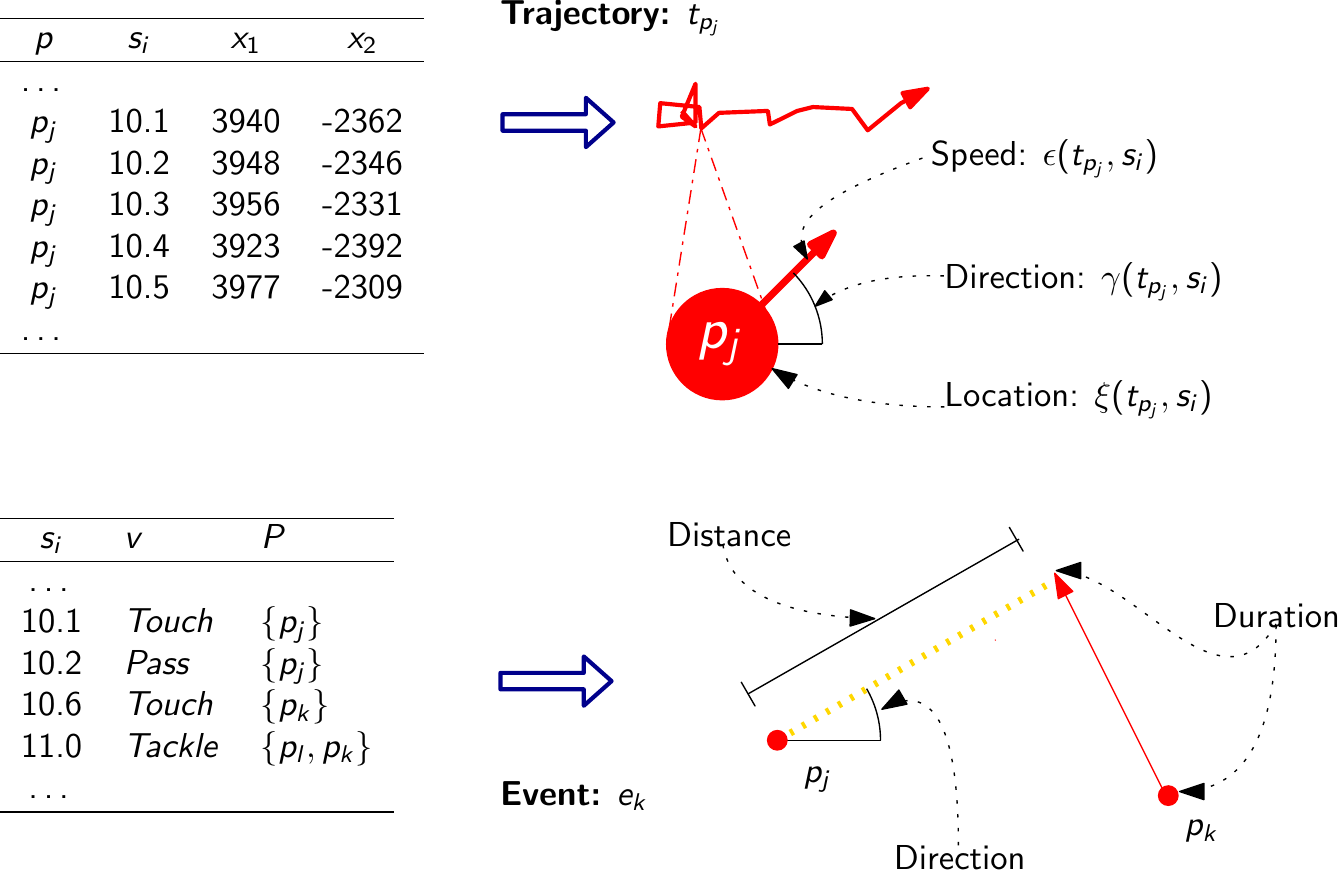}
\end{center}
\caption{Example of the trajectory and event input data and an illustration of their geometric representations. Each trajectory is a sequence of location points, and these can be used to extrapolate the basic geometry of a player at a given time-step. Similarly, the geometry of events such as the \emph{pass} shown, can be computed from the trajectories of the involved players.}
\label{fig:input-data-schematic}
\end{figure}

The movement of players or the ball around the playing area are sampled as a timestamped sequence of location points in the plane, see Fig.~\ref{fig:input-data-schematic}.
The trajectories are captured using optical- or device-tracking and processing systems. 
\emph{Optical tracking systems} use fixed cameras to capture the player movement, and the images are then processed to compute the trajectories~\cite{bradley-2007}.
There are several commercial vendors who supply tracking services to professional sports teams and leagues \cite{tracab,impire,prozone,sportvu}.
On the other hand, \emph{device tracking systems} rely on devices that infer their location, and are attached to the players' clothing or embedded in the ball  or puck. 
These systems can be based on GPS~\cite{catapult} or RFID~\cite{sportvision} technology.

The trajectories produced by these systems are dense, in the sense that the location points samples are uniform and frequent -- in the range of \SIrange[range-units=single]{10}{30}{\hertz}.


The availability of spatio-temporal data for research varies. 
Some leagues capture data from all matches, such as the NBA~\cite{nba-stats} and the German Football Leagues~\cite{impire}, in other cases, teams capture data at their stadia only.
League-wide datasets are not simply larger, but also allow for experiments that control for external factors such as weather, injuries to players, and playing at home and on the road.

\subsection{Event Logs} 

Event logs are a sequence of significant events that occur during a match. 
Events can be broadly categorised as \emph{player events} such as passes and shots; and \emph{technical events}, for example fouls, time-outs, and start/end of period, see Fig.~\ref{fig:input-data-schematic}. 
Event logs may be derived in part from the trajectories of the players involved, however they may also be captured directly from video analysis, for example \citeN{opta} uses this approach. 
This is often the case in sports where there are practical difficulties in capturing player trajectories, such as rugby and American football.

Event logs are qualitatively different from the player trajectories in that they are not dense -- samples are only captured when an event occurs -- however they can be semantically richer as they include details like the type of event and the players involved.

The models and techniques described in the following sections all use object trajectories and/or event logs as their primary input.

\section{Playing Area Subdivision}
\label{sec:subdivision}

Player trajectories and event logs are both low-level representations, and can be challenging to work with. 
      One way 
to deal with this issue is to discretize the playing area into regions and assign the location points contained in the trajectory or event log to a discretized region. 
The frequency -- or intensity -- of events occurring in each region is a spatial summary of the underlying process, alternatively, the playing area may be subdivided into regions such that each region is dominated in some sense by a single player, for example by the player being able to reach all points in the region before any other player.
There are a variety of techniques for producing playing area subdivisions that have been used in the research surveyed here, and are summarised in this section.

\subsection{Intensity Matrices and Maps}
\label{sub:intensity-maps}

\begin{figure}
    \subfloat[Left-back]{
        \includegraphics[width=0.5\linewidth]{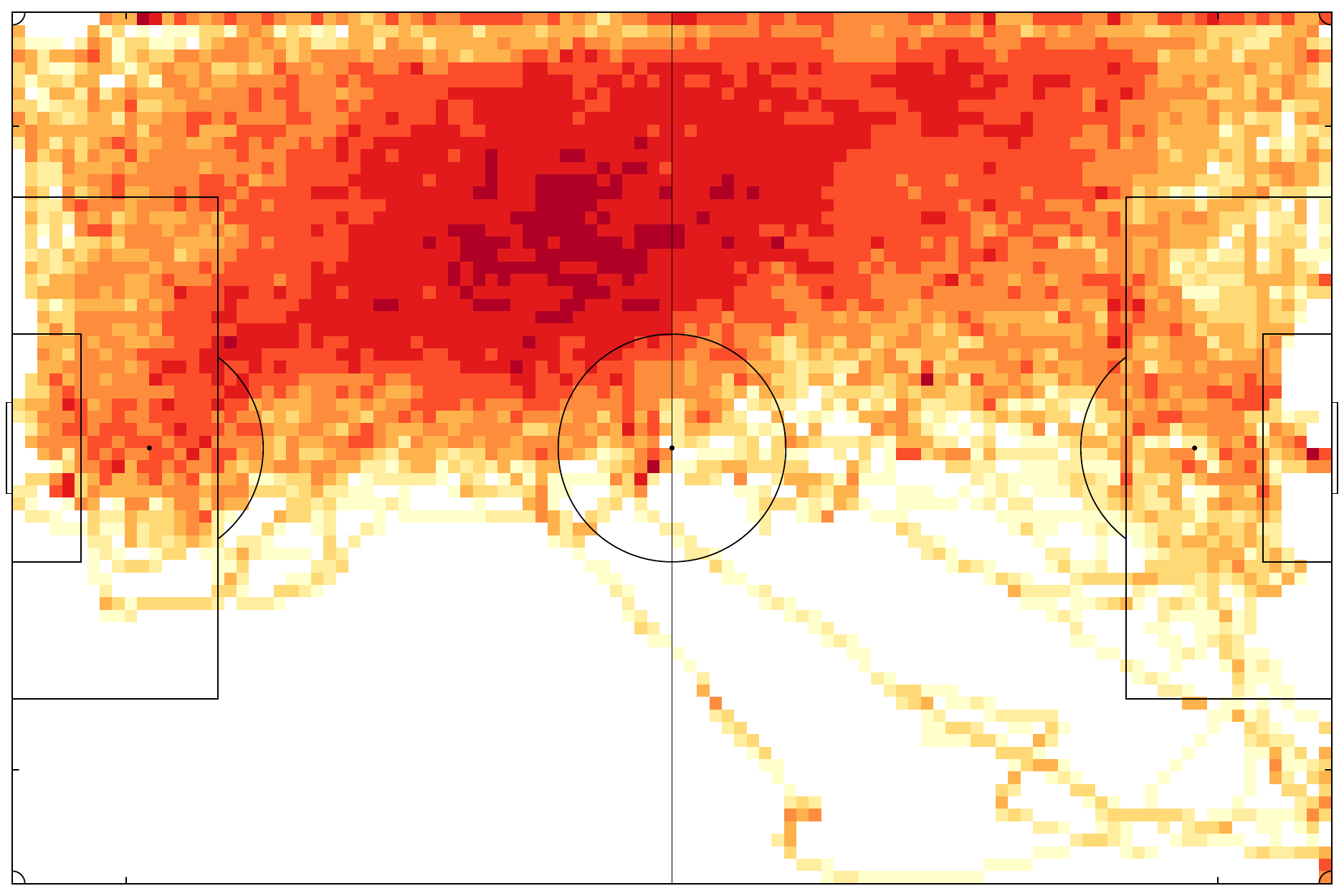}
        \label{fig:football-player-heatmap-108}}
    \subfloat[Striker]{
        \includegraphics[width=0.5\linewidth]{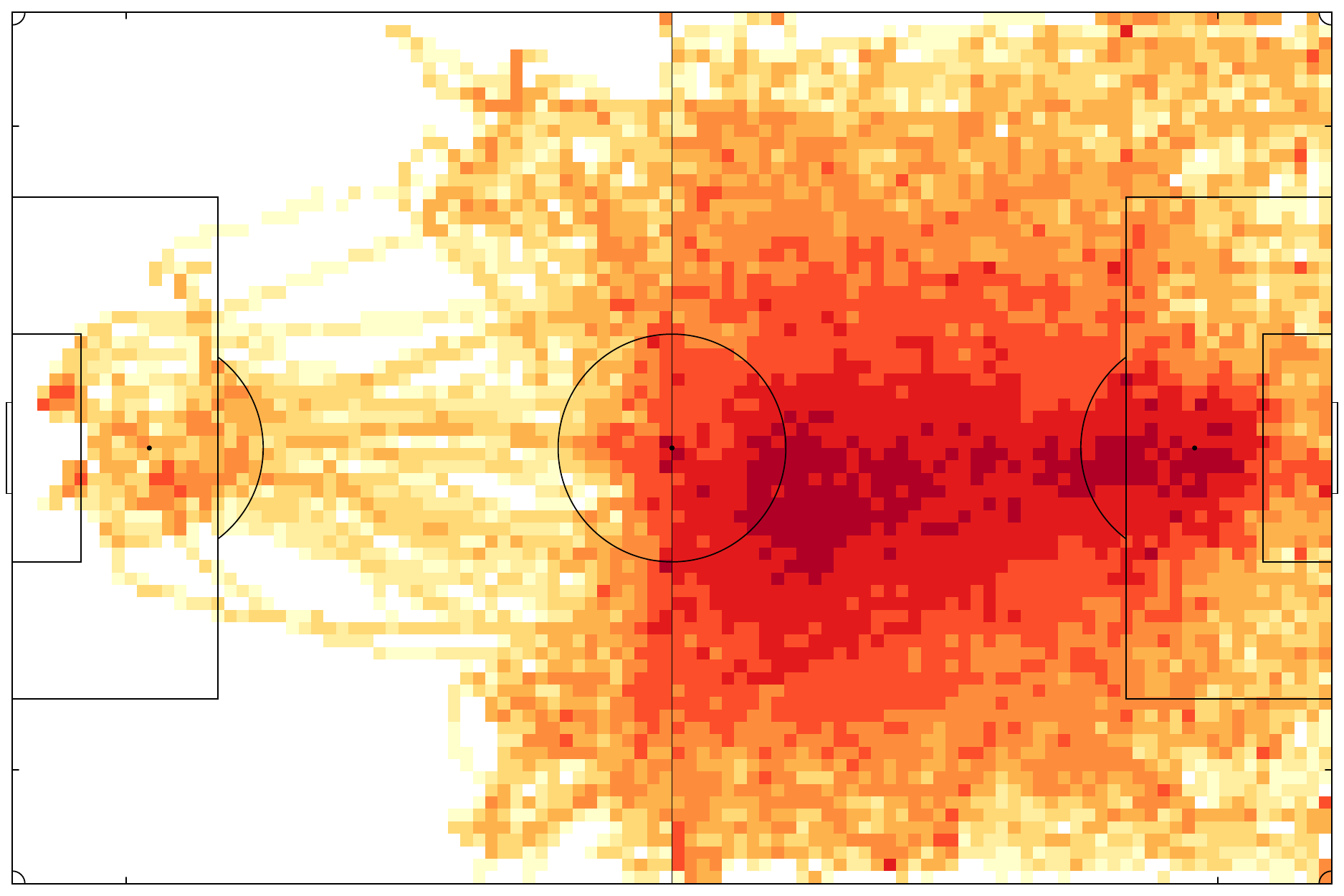}
        \label{fig:football-player-heatmap-2}}
    \caption{Example intensity maps showing areas of the football pitch that the player's occupy. The player trajectories have been oriented such that the play is from left to right. \protect\subref{fig:football-player-heatmap-108} The left-back is positioned on the left of the field, but is responsible for taking attacking corner-kicks from the right. 
        \protect\subref{fig:football-player-heatmap-2} The striker predominantly stays forward of the half-way line, however will retreat to help defend corner-kicks.}
    \label{fig:intensity-map}
\end{figure}

Spatial data from team sports have the useful property that they are constrained to a relatively small and symmetric playing area -- the pitch, field or court. 
The playing area may be subdivided into regions and events occurring in each region can be counted to produce an intensity matrix, and can be visualised with an intensity map, see Fig.~\ref{fig:intensity-map}.
This is a common preprocessing step for many of the techniques described in subsequent sections. 

\begin{figure}
    \centering
    \subfloat[Hand-designed]{
        \centering
        \includegraphics[width=0.5\linewidth]{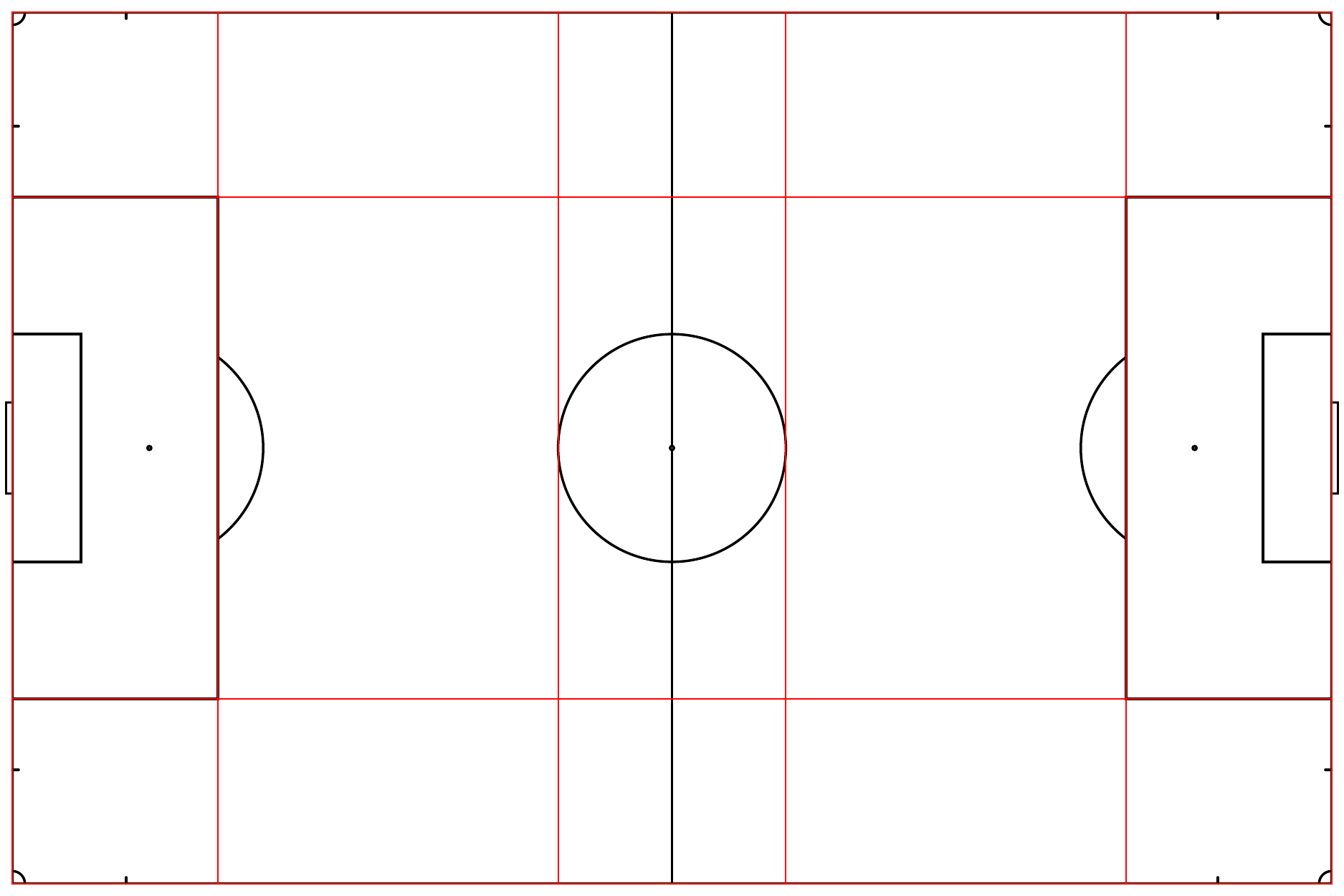}
        \label{fig:camerino-subdiv}}
    \\
    \subfloat[Hand-designed]{
        \includegraphics[width=0.32\linewidth]{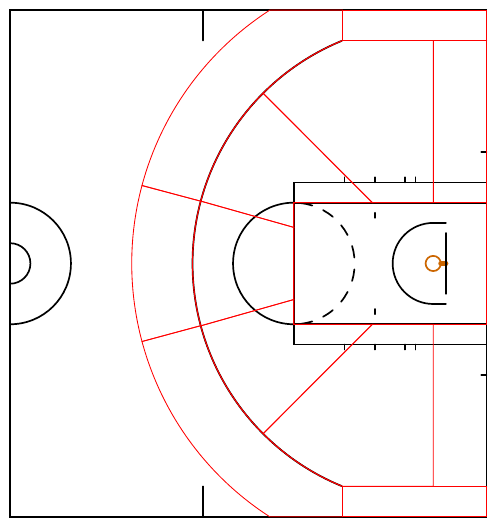}
        \label{fig:maheswaran-subdiv}}
    \subfloat[Cartesian grid]{
        \includegraphics[width=0.32\linewidth]{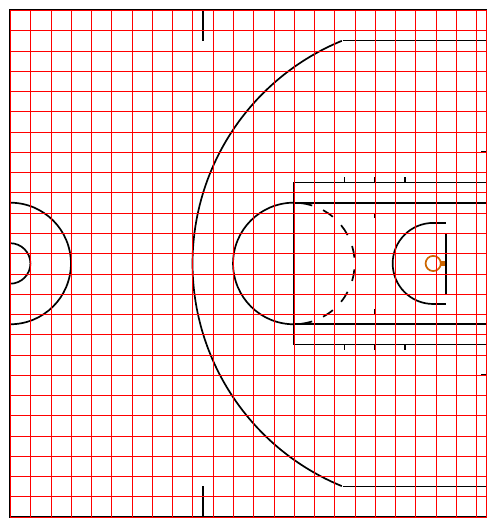}
        \label{fig:cartesian-subdiv}}
    \subfloat[Polar grid]{
        \includegraphics[width=0.32\linewidth]{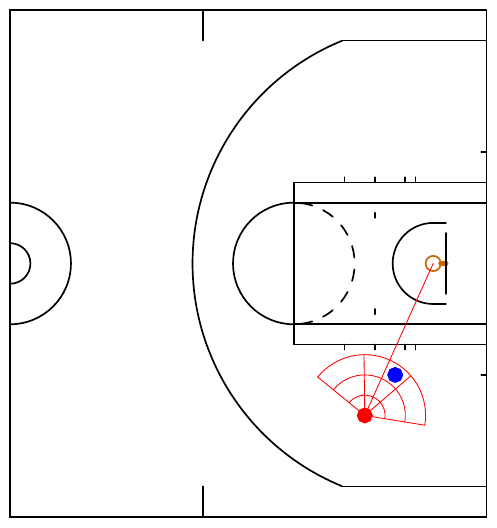}
        \label{fig:yue-polar-subdiv}}
    \caption{Examples of subdivisions used to discretize locations:  
        \protect\subref{fig:camerino-subdiv}, \protect\subref{fig:maheswaran-subdiv} hand-designed subdivision reflecting expert knowledge of game-play in basketball~\protect\cite{maheswaran-2014} and football~\protect\cite{camerino-2012}; 
        \protect\subref{fig:cartesian-subdiv} subdivision of court into unit-squares~\protect\cite{cervone-2014a};
        \protect\subref{fig:yue-polar-subdiv} polar subdivision where origin is centred on ball-carrier and grid is aligned with the basket~\protect\cite{yue-2014}.}
    \label{fig:basketball-subdivisions}
\end{figure}

When designing a spatial discretization, the number and shape of the induced regions can vary.
A common approach is to subdivide the playing area into rectangles of equal size~\cite{bialkowski-2014a,borrie-2002,cervone-2014a,franks-2015,lucey-2012,Narizuka2014a,shortridge-2014}, for example see Fig.~\ref{fig:cartesian-subdiv}. 
However, the behaviour of players may not vary smoothly in some areas.
For example: around the three-point line on the basketball court, a player's propensity to shoot varies abruptly; or the willingness of a football defender to attempt a tackle will change depending on whether they are inside the penalty box. 
The playing area may be subdivided to respect such predefined assumptions of the player's behaviour. 
\citeN{camerino-2012} subdivides the football pitch into areas that are aligned with the penalty box, see Fig.~\ref{fig:camerino-subdiv}, and interactions occurring in each region where counted.
Similarly, \citeN{maheswaran-2014} and \citeN{goldsberry-2013} define subdivisions of the basketball half-court that conforms with the three-point line and is informed by intuition of shooting behaviour, see Fig.~\ref{fig:maheswaran-subdiv}.

Transforming the playing area into polar space and inducing the subdivision in that space is an approach used in several papers. 
This approach reflects the fact that player behaviour may be similar for locations that are equidistant from the goal or basket.
Using the basket as the origin, polar-space subdivisions were used by \citeN{reich-2006} and by \citeN{maheswaran-2012}. 
\citeN{yue-2014} used a polar-space subdivision to discretize the position of the players marking an attacking player. 
Under this scheme, the location of the attacking player was used as the origin, and the polar space aligned such that the direction of the basket is at \SI{0}{\degree}, see Fig.~\ref{fig:yue-polar-subdiv}.

Given a subdivision of the playing area, counting the number of events by each player in each region induces a discrete spatial distribution of players' locations during the match.
This can be represented as an $\mathbb{R}_{+}^{N \times K}$ intensity matrix containing the counts $X$ for $N$ players in each of the $R$ regions of the subdivision. 
The event $X$ may be the number of visits by a player to the region, e.g. \citeN{maheswaran-2014} used the location points from player trajectories to determine whether a cell was visited. 
\citeN{bialkowski-2014} used event data such as passes and touches made by football players to determine the regions a player had visited.

The number of passes or shots at goal that occur in each region may also be counted. 
For example, many papers counted shots made in each region of a subdivision of a basketball court \cite{franks-2015,goldsberry-2013,maheswaran-2012,reich-2006,shortridge-2014}.
Similarly, \citeN{borrie-2002}, \citeN{camerino-2012}, \citeN{Narizuka2014a}, and \citeN{cervone-2014a} counted the number of passes made in each region of a subdivision of the playing area.

\subsection{Low-rank Factor Matrices}
\label{sub:matrix-factorization}

Matrix factorization can be applied to intensity matrices described in Section~\ref{sub:intensity-maps}, to produce a compact, low-rank representation.
This approach has been used in several papers to model shooting behaviour in basketball~\cite{cervone-2014a,franks-2015,yue-2014}.
The insight that motivates this technique is that similar types of players tend to shoot from similar locations, and so each player's shooting style can be 
      modelled as a combination of a few 
distinct \emph{types}, where each \emph{type} maps to a coherent area of the court that the players are likely to shoot from.

\begin{figure}
    \centering
    \subfloat[Corner three-point] {
        \includegraphics[width=0.32\linewidth]{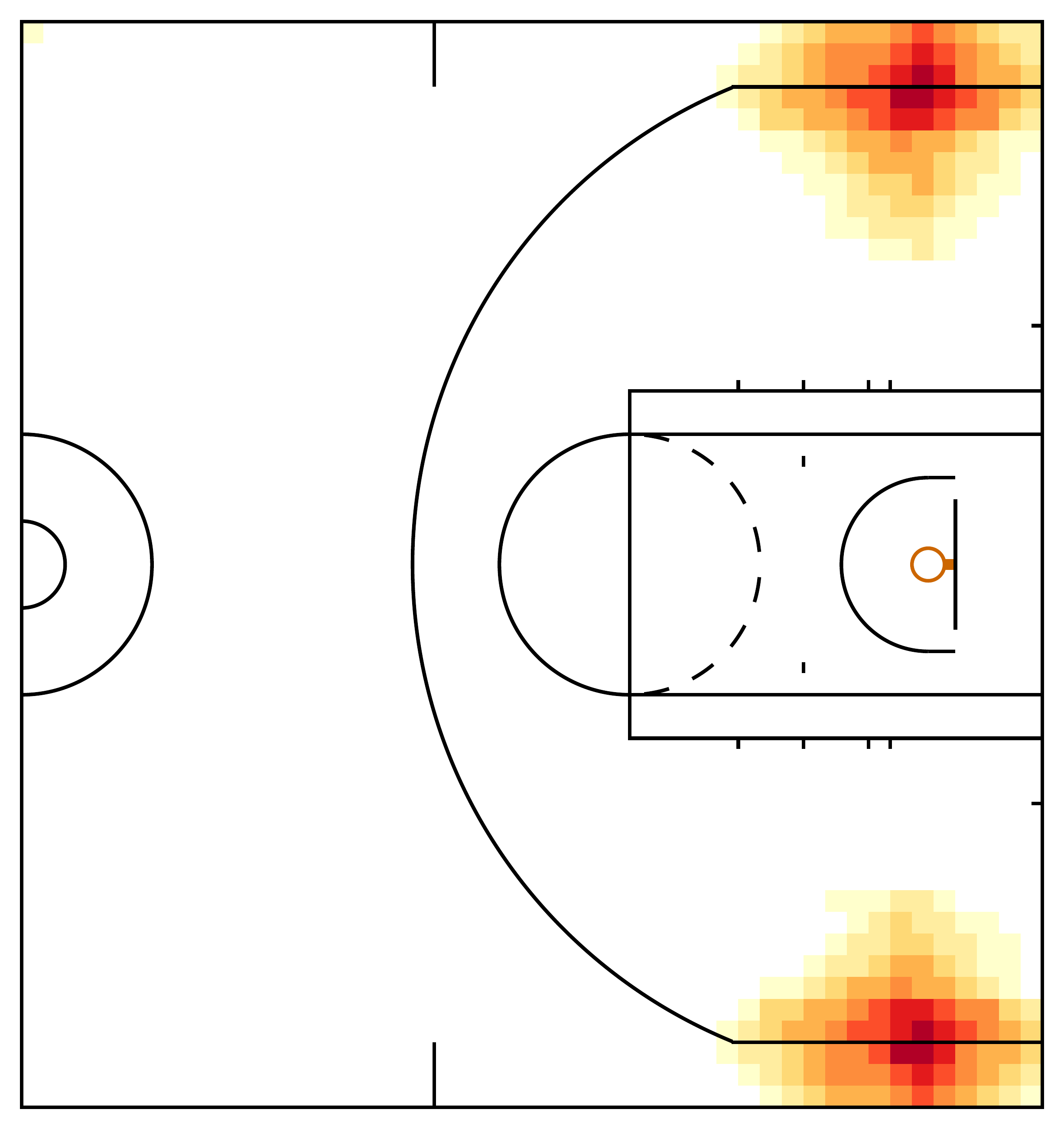}
        \label{fig:miller-heatmap-0}}
    \subfloat[Top-of-key three-point] {
        \includegraphics[width=0.32\linewidth]{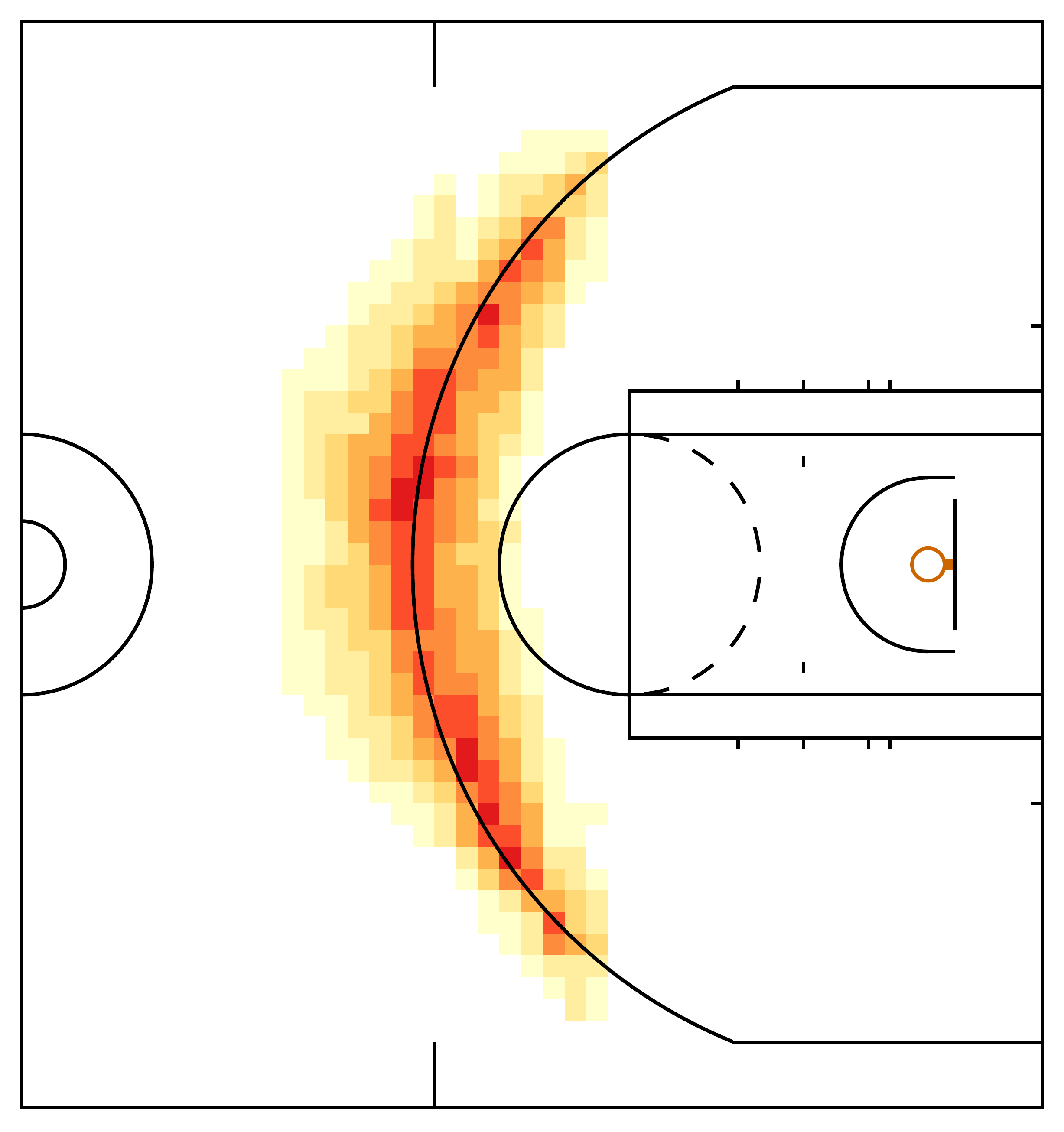}
        \label{fig:miller-heatmap-1}}
    \subfloat[Right low-post] {
        \includegraphics[width=0.32\linewidth]{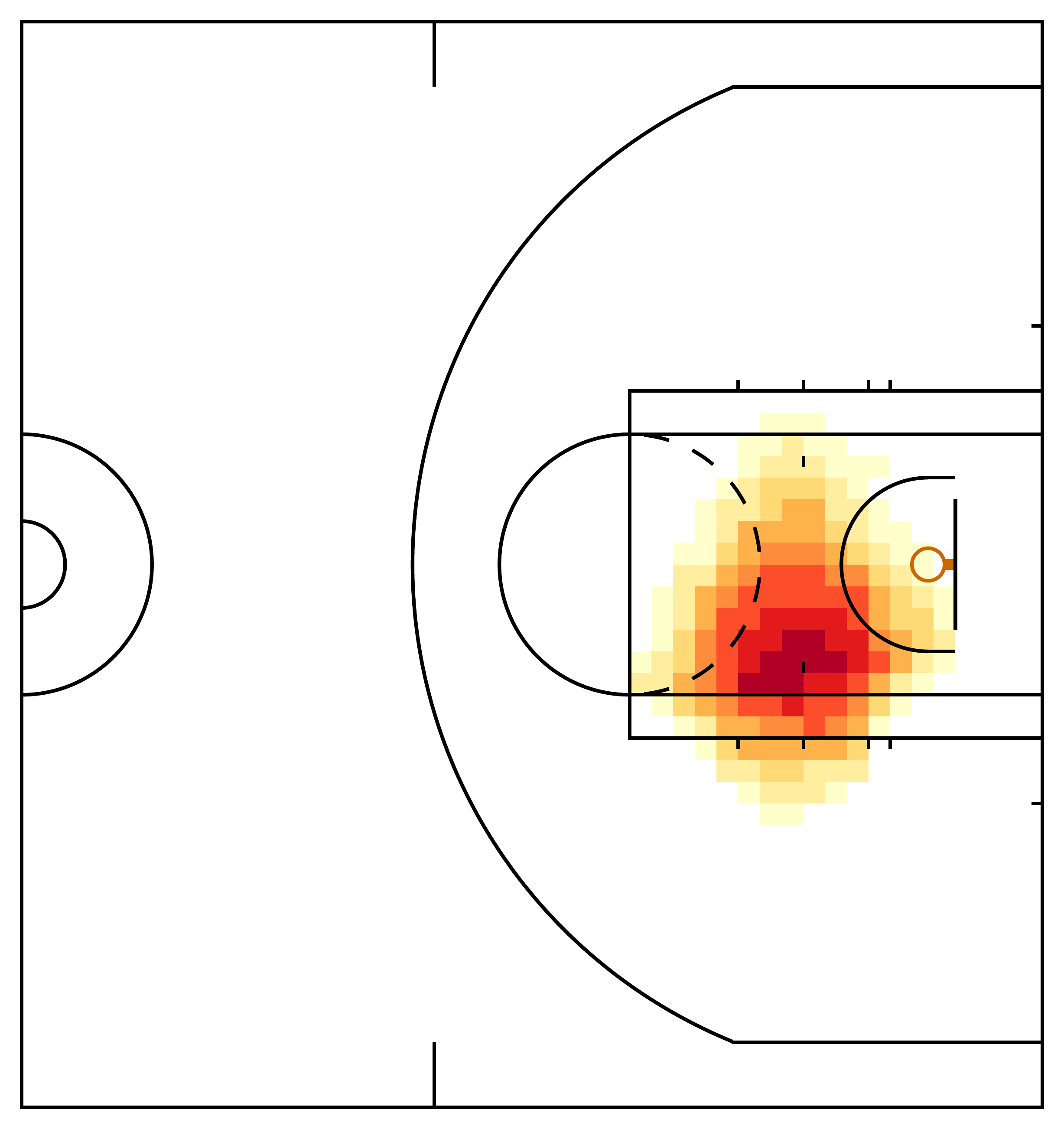}
        \label{fig:miller-heatmap-2}}
\caption{Examples of spatial bases induced by using non-negative matrix factorization. Each basis represents an intensity map of where a subset of players tend to shoot from. 
    Shown are three spatial basis intensity maps that represent defined shooting locations.
}
\label{fig:miller-heatmap}
\end{figure}

The input is an intensity matrix $X \in \mathbb{R}^{N \times V}$. Two new matrices $W \in \mathbb{R}_{+}^{N \times K}$ and $B \in \mathbb{R}_{+}^{K \times V}$ are computed such that $WB \approx X$ and $K \ll N, V$. 
The $K$ spatial bases in $B$ represent areas of similar shooting intensity, and the $N$ players' shooting habits are modelled as a linear combination of the spatial bases.
The factorization is computed from $X$ by minimizing some distance measure between $X$ and $WB$, under the constraint that $W$ and $B$ are non-negative. 
The non-negativity constraint, along with the choice of distance function encourages sparsity in the learned matrices. 
This leads to intuitive results: each spatial basis corresponds to a small number of regions of the halfcourt; and 
      the shooting style of each player is modelled as the mixture
of a small number of bases, see Fig.~\ref{fig:miller-heatmap} for examples of learned spatial bases.

\citeN{miller-2014} used non-negative matrix factorization to represent shooting locations in basketball. They observe that the shooting intensity should vary smoothly over the court space, and thus fit a Log-Gaussian Cox Process to infer a smooth intensity surface 
over the intensity matrix, which is then factorized.

\citeN{yue-2014} used non-negative matrix factorization to model several event types: shooting; passing and receiving. 
They include a spatial regularization term in the distance function used when computing the matrix factorization, and claim that spatial regularization can be seen as a frequentist analog of the Bayesian Log-Gaussian Cox process used by \citeN{miller-2014}.

\citeN{cervone-2014a} also used non-negative matrix factorization to find a basis representing player roles, based on their occupancy in areas of the court. 
Players who are similar to a given player were identified as those who are closest in this basis, and this was used to compute a similarity matrix between players.

\subsection{Movement Models and Dominant Regions}
\label{sec:dominant-region}

A team's ability to control space is considered a key factor in the team's performance, and was one of the first research areas in which computational tools were developed. Intuitively a player dominates an area if he can reach every point in that area before anyone else (see Definition~\ref{def:DR}). An early algorithmic attempt to develop a computational tool for this type of analysis was presented by \citeN{taki-1996}, which defined the \emph{Minimum Moving Time Pattern} -- subsequently renamed the \emph{Motion Model} -- and the \emph{Dominant Region}. 

\subsubsection{Motion Model} \label{sssec:MM}

The motion model presented by \citeN{taki-1996} is simple and intuitive: it is a linear interpolation of the acceleration model.
It assumes that potential acceleration is the same in all directions when the player is standing still or moving very slowly. As speed increases it becomes more difficult to accelerate in the direction of the movement. 
However, their model did not account for deceleration and hence is only accurate over short distances. 


\citeN{Fujimura2005} presented a more realistic motion model, in particular they incorporated a resistive force that decrease the acceleration. 
The maximum speed of a player is bounded, and based on this assumption, \citeN{Fujimura2005} formulated the following equation of motion:
\begin{equation} \label{eq:MM}
 m \frac{d}{dt} v = F-kv,
\end{equation}
where $m$ is the mass, $F$ is the maximum driving force, $k$ is the resistive coefficient, and $v$ is the velocity. The solution of the equation is:
$$v = \frac {F}{k} - (\frac {F}{k}-v_0) \cdot \exp (-\frac {k}{m} t),$$
where $v_0$ is the velocity at time $t=0$. If the maximum speed $v_{\max}=F/k$ and the magnitude of the resistance $\alpha=k/m$ are known, then the motion model is fixed. To obtain $\alpha$ and $v_{\max}$, \citeN{Fujimura2005} studied players' movement on video and empirically estimated $\alpha$ to be $1.3$ and $v_{\max}$ as $7.8$m/s.
This is then generalised to two dimensions as follows:
 $$m \frac{d}{dt} \vec{v} = \vec{F} -k \vec{v}.$$
Solving the equation we get that all the points reachable by a player, starting at position $x_0$ with velocity $\vec{v_0}$, can reach point $x$ within time $t$ form the circular region centred at 
$$x_0+\frac{1-e^{-\alpha t}}{\alpha} \cdot \vec{v_0} \quad \text{with radius}
\quad \vec{v}_{\max} \cdot \frac{1-e^{-\alpha t}}{\alpha}.$$
They compared this model empirically and observed that the model yields a good approximation of actual human movement, but they stated that a detailed analysis is a topic for future research.

A different model was used in a recent paper by \citeN{cervone-2014a} with the aim to predict player movement in basketball. They present what they call a micro-transition model. The micro-transition model describes the player movement during a single possession of the ball. 
Separate models are then used for defense and attack. Let the location of an attacking player $\ell$ at time $t$ be $(x^{\ell}(t),y^{\ell}(t))$. Next they model the movement in the $x$ and $y$ coordinates at time $(t+\eps)$ using
\begin{equation} \label{eqn:motion}
  x^{\ell}(t+\eps)=x^{\ell}(t)+\alpha^{\ell}_x[x^{\ell}(t)-x^{\ell}(t-\eps)]+ \eta_x^{\ell}(t)],
\end{equation}
and analogously for $y^{\ell}(t+\eps)$.
This expression derives from a Taylor series expansion of the function for determining the ball-carrier's position such that $\alpha^{\ell}_x[x^{\ell}(t)-x^{\ell}(t-\eps)] \approx \eps x^{\ell}(t)$, and $\eta^{\ell}_x(t)$ represents the contribution of higher order derivatives modelling accelerations and jerks. 
When a player receives the ball outside the three-point line, the most common movement is to accelerate towards the basket.
On the other hand, a player will decelerate when closer to the basket. 
Players will also accelerate away from the boundary of the court as they approach it. To capture this behaviour the authors suggest mapping a player's location to the additive term $\eta_x^{\ell}(t)$ in~\eqref{eqn:motion}. 
The position of the five defenders are easier to model, conditioned on the evolution of the attack's positions, see \citeN{cervone-2014a} for details.

Next we consider how the motion models have been used to develop other tools.

\subsubsection{Dominant Regions} \label{sssec:DR}

\begin{figure}
\begin{center}
 \includegraphics[width=.8\textwidth]{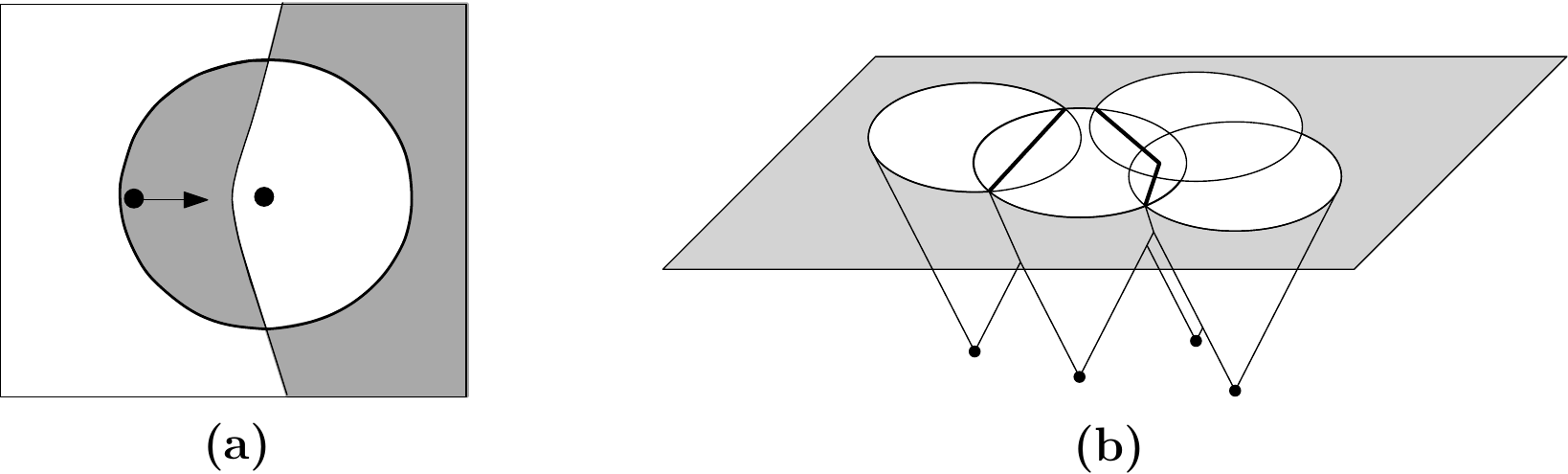}
\end{center}
\caption{(a) Showing the dominant region for two players. The left player is moving to the right with high speed and the right player is standing still. Using the motion models discussed in Section~\ref{sssec:MM} the resulting dominant region for a single player might not be connected. (b) A standard approach used in computational geometry to subdivide the plane is to compute the projection of the intersection of surfaces in three dimensions onto the plane.}
\label{fig:MM}
\end{figure}

The original paper by~\citeN{taki-1996} defined the dominant region as:
\begin{definition} \label{def:DR}
 The \emph{dominant region} of a player $p$ is the region of the playing area where $p$ can arrive before any other player.
\end{definition}
Consequently the subdivision induced by the dominant regions for all players will partition the playing area into cells. 
In a very simple model where acceleration is not considered, the dominant region is equivalent to the Voronoi region and the subdivision can be efficiently computed~\cite{fortune-1987}. However, for more elaborate motion models, such as the ones described in Section~\ref{sssec:MM}, the distance function is more complex. 
      For some motion models the dominant region may not be a connected area~\cite{taki-2000}, 
an example is shown in Fig.~\ref{fig:MM}a. A standard approach used to compute the subdivision for a complex distance function is to compute the intersection of surfaces in three dimensions, as shown in Fig.~\ref{fig:MM}b. However, this is a complex task and time-consuming for non-trivial motion models. Instead approximation algorithms have been considered in the literature.

Taki and Hasegawa~\shortcite{taki1999,taki-2000} implemented algorithms to compute dominant regions, albeit using a simple motion model. Instead of computing the exact subdivision they considered the $640\times480$ pixels that at that time formed a computer screen and for each pixel they computed the player that could reach that pixel first, hence, visualizing the dominant regions. The same algorithm for computing the dominant region was used by~\citeN{Fujimura2005}, although they used a more realistic motion model, see Section~\ref{sssec:MM}.

However, the above algorithms were shown to be slow in practice, for example preliminary experiments by \citeN{nakanishi-2009} stated that the computation requires \SIrange{10}{40}{\second} for a $610\times420$ grid. To achieve the real-time computation required for application in the RoboCup robot football tournament~\cite{kitano-1997}, the authors proposed an alternative approach. Instead of computing the time 
      required for every player to get to every point,
\citeN{nakanishi-2009} used a so-called \emph{reachable polygonal region} (RPR). The RPR of a player $p$ given time $t$ is the region that $p$ can reach within time $t$. An advantage with using the RPR for computing dominant regions is that more complex motion models can be used by simply drawing the RPR for different values of $t$. They presented the following high-level algorithm. Given a sequence of time-steps $t_i$, $1\leq i \leq k$ compute the RPRs for each player and each time-step. 
The algorithm then iterates through the sequence of time-steps and for each pair of players, the \emph{partial dominant regions} are constructed from the RPRs.
The partial dominant regions are then combined with the dominant regions computed in the previous time-step to form new dominant regions. 
Assuming that the RPR is a convex area for any $p$ and any $t$, Nakanishi~\etal claim a factor of $1000$ improvement in computation time at the cost of roughly a $10\%$ drop in accuracy.

\begin{figure}
    \subfloat[~]{
        \includegraphics[width=0.40\linewidth]{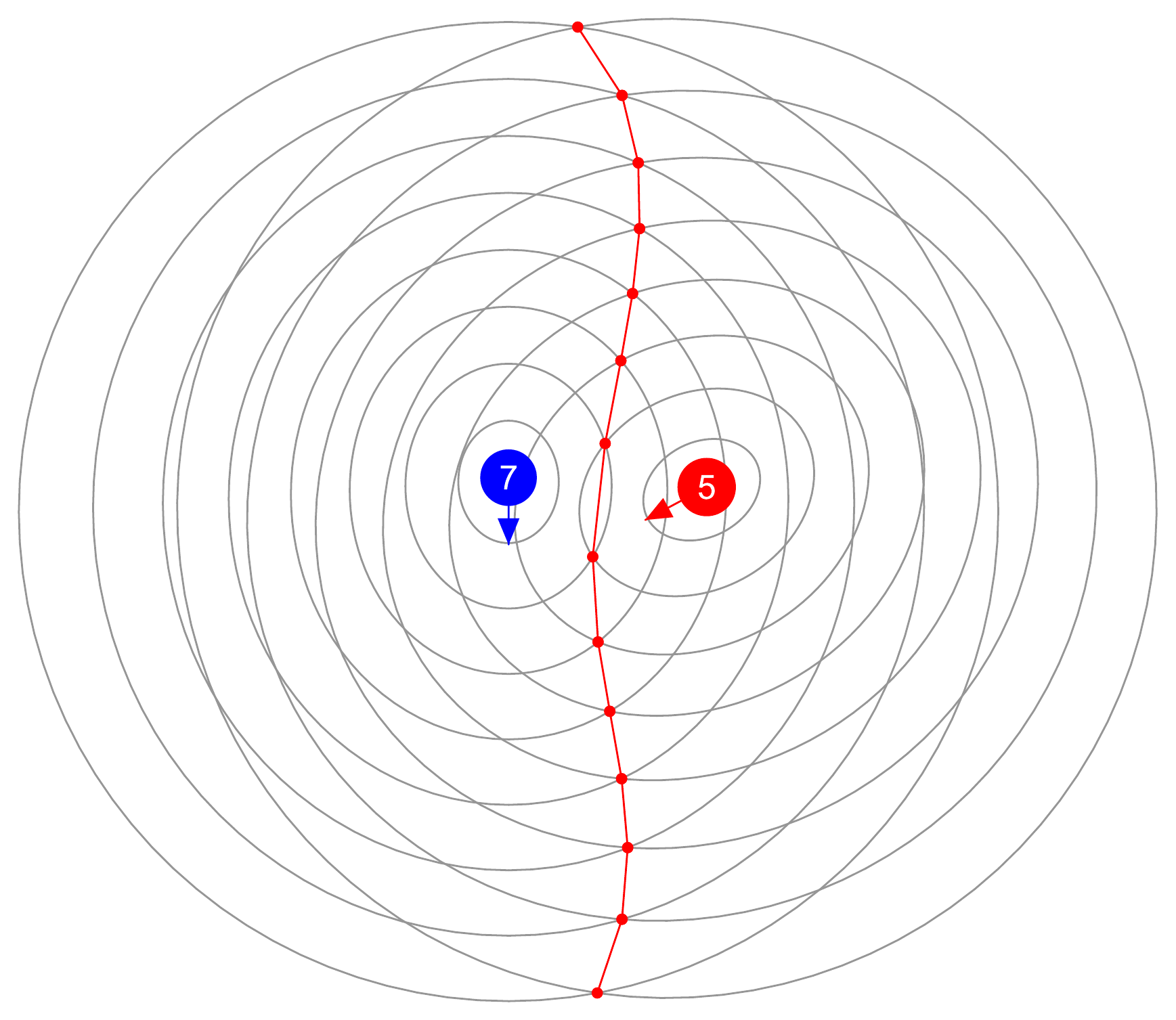}
        \label{fig:DiscBis:bisector}}
    \subfloat[~]{
        \includegraphics[width=0.59\linewidth]{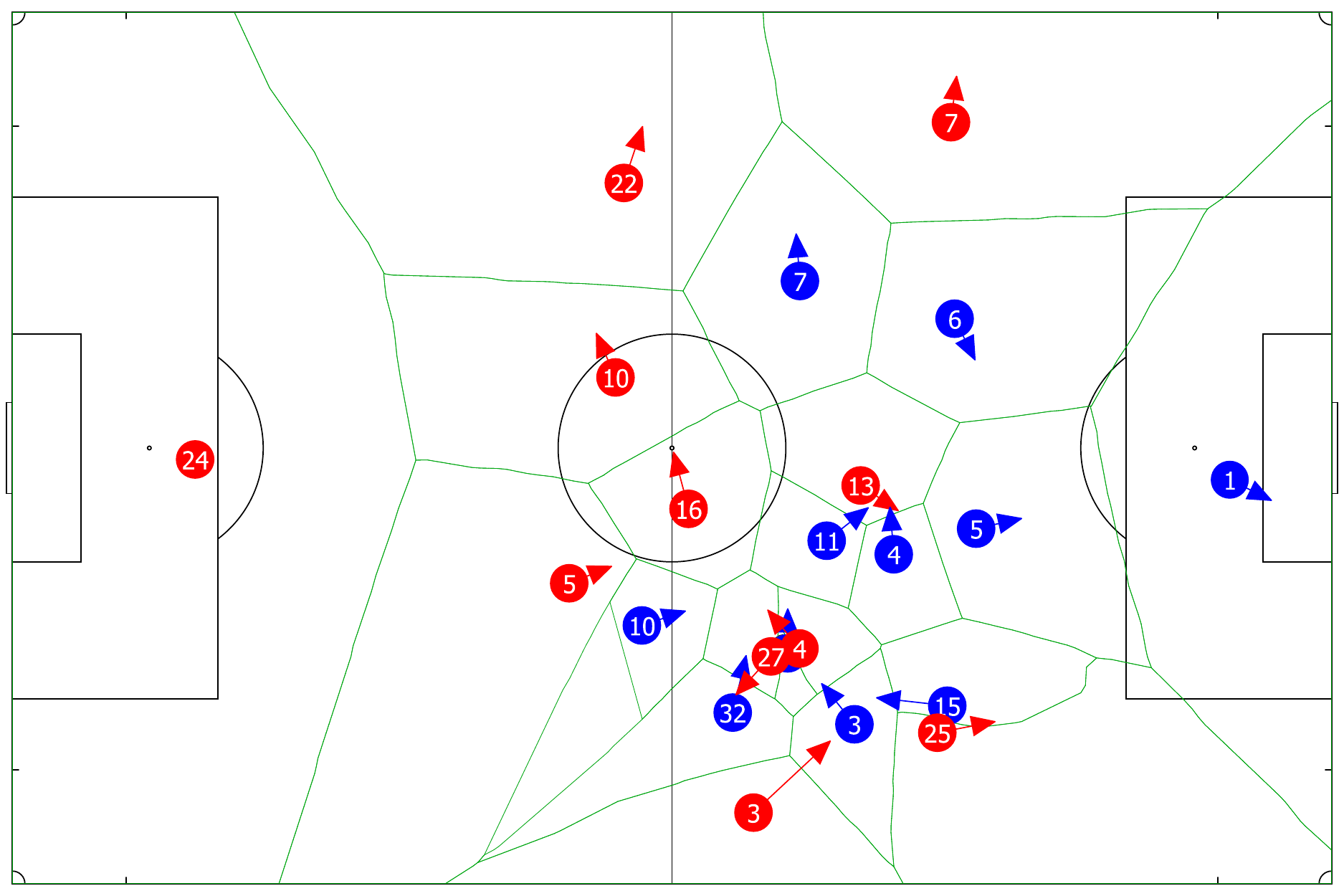}
        \label{fig:DiscBis:dominant-region}}
    \caption{\protect\subref{fig:DiscBis:bisector} An approximate bisector between two players using the intersection points of the RPRs. \protect\subref{fig:DiscBis:dominant-region} An example of the approximate dominant region subdivision by \protect\citeN{gudmundsson-2014}.}
    \label{fig:DiscBis}
\end{figure}

\citeN{gudmundsson-2014} used RPRs induced from real trajectory data. They also presented an algorithm for constructing an approximate dominant region subdivision, which is superficially similar to the algorithm by \cite{nakanishi-2009}. However, instead of computing partial dominant regions for each pair of players at each time-step, an approximate bisector is constructed for every pair of players. An example of an approximate bisector between two players is shown in Fig.~\ref{fig:DiscBis}a, and in Fig.~\ref{fig:DiscBis}b the final subdivision generated by the algorithm in \citeN{gudmundsson-2014} is depicted.  

A closer study of a player's dominant region was performed by~\citeN{fonseca-2012} in an attempt to
      describe the spatial interaction between players.
They considered two variables denoting the 
      smallest distance between two teammates and the size of the dominant region.
They observed that the individual dominant regions 
      seem to be larger for the attacking team.
      They also found that for the defending team the two measures were more irregular which indicates that their movement was more unpredictable that the movement of the attacking team.

According to the authors, the player and team dominant regions 
      detect certain match events
such as ``when the ball is received by an attacker inside the defensive structure, revealing behavioural patterns that may be used to explain the performance outcome."

\citeN{ueda2014} compared the team-area and the dominant region (within the team-area) during offensive and defensive phases. 
The \emph{team area} is defined as the smallest enclosing orthogonal box containing all the field players of the defending team. 
      The results seem to show that there exists a correlation between the ratio of the dominant region to team area, and the performance of the team's offence and defence.
Dominant regions of successful attacks were thinner than those for unsuccessful attacks that broke down with a turnover event located near the centre of the playing area. 
      The conclusion was that the dominant region is closely connected to the offensive performance, hence, perhaps it is possible to evaluate the performance of a group of players using the dominant region.
\begin{open}
    The function modelling player motion used in dominant region computations has often been simple for reasons of tractability or convenience. 
Factors such as the physiological constraints of the players and \emph{a priori} momentum have been ignored.
A motion function that faithfully models player movement and is tractable for computation is an open problem.
\end{open}


\subsubsection{Further Applications}
The dominant region is a fundamental structure that has been shown to support several other interesting measures, and are discussed next.

\begin{figure}
\begin{center}
 \includegraphics[width=0.7\textwidth]{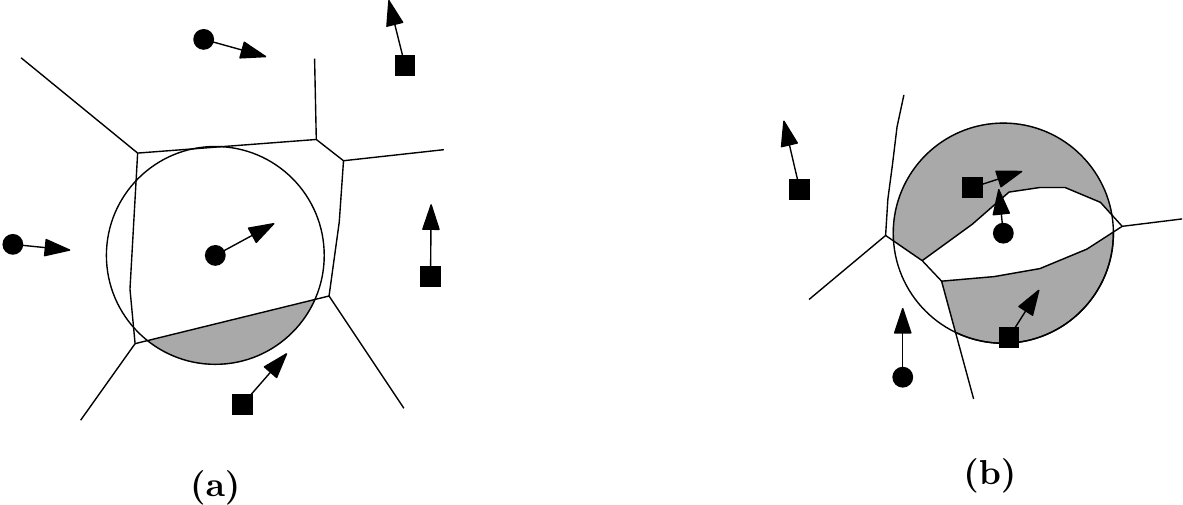}
 \caption{Comparing the pressure that the encircled player is under in the two pictures shows that the encircled player in the right figure is under much more pressure.}
  \label{fig:Applications}
\end{center}
\end{figure}

\begin{enumerate}
 \item {\bf (Weighted) Area of team dominant region.} \citeN{taki-1996} defined a \emph{team dominant region} as the union of dominant regions of all the players in the team. 
     Variations in the size of the team dominant region was initially regarded by \cite{taki-1996} as a strong indication on the performance of the team. 
     However, \citeN{Fujimura2005} argued that the size of a dominant region does not capture the contribution of a player. Instead they proposed using a \emph{weighted} dominant region, by either weighting with respect to the distance to the goal, or with respect to the distance to the ball.
     They argued that both these approaches better 
           model the contribution of a player compared to simply using the size of the dominant region.
     However, no further analysis was performed.

    \citeN{Fujimura2005} also suggested that the weighted area of dominant regions can be used to evaluate attacking teamwork: tracking the weighted dominant region (``defensive power") over time for the defender marking each attacker will indicate the attacker's contribution to the team.

 \item {\bf Passing evaluation.} 
          A player's \emph{passable area} is the region of the playing area where the player can potentially receive a pass. The size and the shape of the passable area depends on the motion model, and the positions of the ball and the other players. Clearly this is also closely related to the notion of dominant region.     
    \begin{definition} \cite{gudmundsson-2014}
       A player $p$ is open for a pass if there is some direction and (reasonable) speed that the ball can be passed, such that $p$ can intercept the ball before all other players.
    \end{definition}


   \citeN{taki-2000} further classified a pass as ``successful" if the first player that can receive the pass is a player from the same team. 
   This model was extended and implemented by \citeN{Fujimura2005}, as follows. They empirically developed a motion model for the ball, following formula~(\ref{eq:MM}) in Section~\ref{sssec:MM}. 
   They then defined the \emph{receivable pass variation} (RPV) for each player to be the number of passes the player can receive among a set of sampled passes. 
   They experimentally sampled \SI{54000} passes by discretizing $[0,2\pi)$ into \SI{360} unit directions and speeds between \SI{1} and \SI{150}{\km\per\hour} into \SI{150} units. 

   \citeN{gudmundsson-2014} also used a discretization approach, but viewed the problem slightly differently. 
   Given the positions, speeds and direction of motion of the players, they approximated who is open for a pass for each discretized ball speed. 
   For each fixed passing speed they built RPRs for each player and the ball over a set of discrete time-steps. Then 
   an approximate bisector is computed between the ball and the player. Combining the approximate bisectors for all the players,
   a piecewise linear function $f$ is generated over the domain $[0,2\pi)$. The segments of the bisectors that lie on the lower envelope of $f$ map to intervals on the domain where the player associated with the bisector is open for a pass. An example of the output is shown in Fig.~\protect\ref{fig:football-passability} for a fixed ball speed.

\begin{figure}
\begin{center}
    \includegraphics[width=0.8\linewidth]{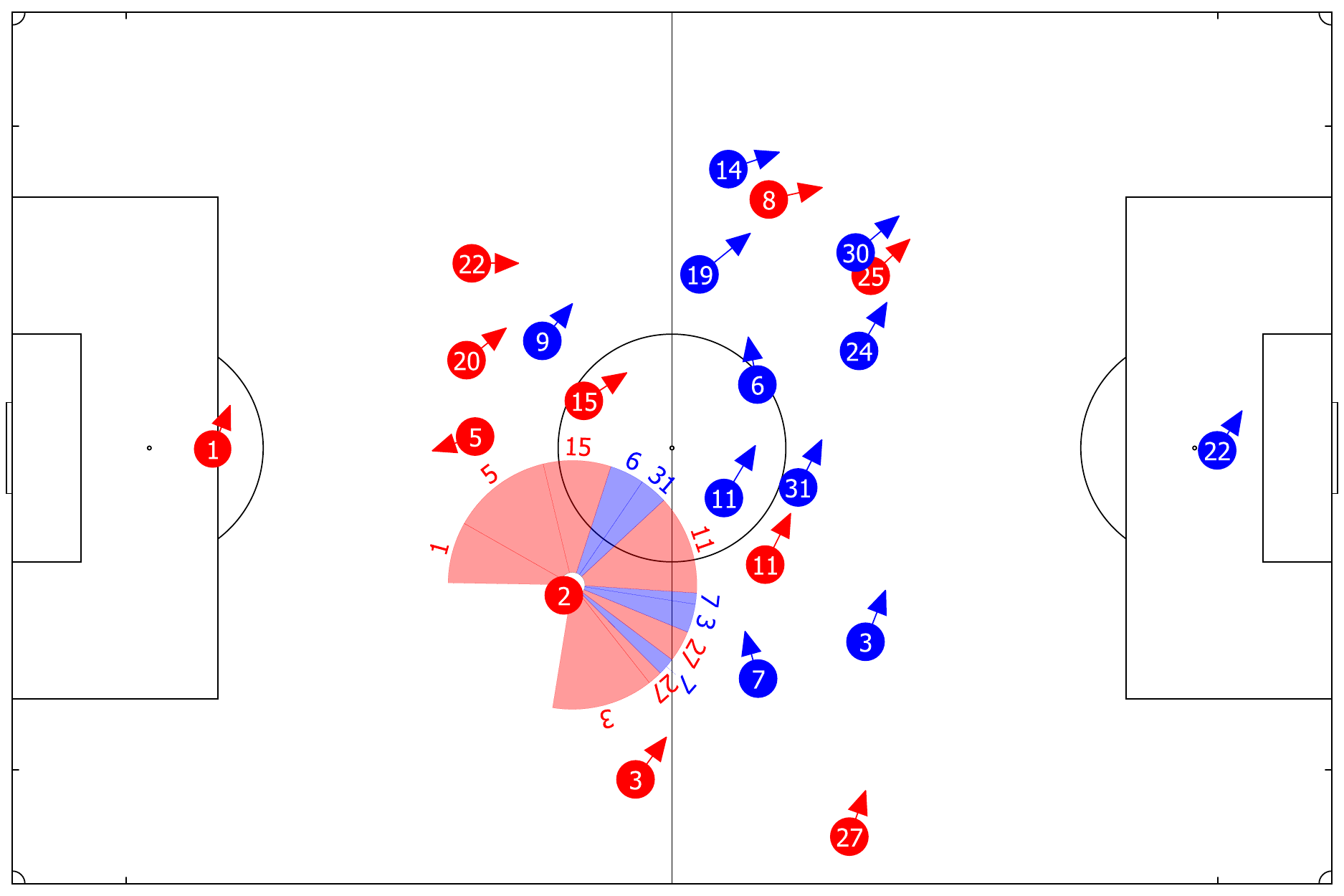}
\end{center}
\caption{Available receivers of a pass by player Red~$2$ where velocity of the ball is \protect\SI{20}{\meter\per\second}. Each sector represents an interval on $[0,2\pi)$ that indicates which player may receive the pass. Players may receive a pass made at more than one interval, for example Blue~$7$.}
\label{fig:football-passability}
\end{figure}

\begin{open}
  The existing tools for determining whether a player is open to receive a pass only consider passes made along the shortest path between passer and receiver and where the ball is moving at constant velocity.
  The development of more realistic model that allows for aerial passes, effects of ball-spin, and variable velocities is an interesting research question.
\end{open}

\item {\bf Spatial Pressure.} An important tactical measure is the amount of spatial pressure the team exerts on the opposition.
    Typically when a team believes that the opponent is weak at retaining possession of the ball, then a high pressure tactic is used. 
    \citeN{taki-1996} defined spatial pressure for a player $p$ as:
    $$m\cdot (1-P) + (1-m)\cdot (1-d/D),$$
    where, for a fixed radius $r$, $P$ denotes the fraction of the disk of radius $r$ with center at $p$ that lies within the dominant region of opposing players, $d$ is the distance between $p$ and the ball, $D$ is the maximal distance between $p$ and any point on the playing area, and $m$ is a preset weight. 
    This definition was also used by \citeN{horton-2015}. See Fig.~\ref{fig:Applications} for two examples of spatial pressure.

    \begin{open}
        The definition of spatial pressure in \citeN{taki-1996} is simple and does not model effects such as the direction the player is facing or the direction of pressuring opponents, both of which would intuitively be factors that ought to be considered.
        Can a model that incorporates these factors be devised and experimentally tested?
    \end{open}

\item {\bf Rebounding.} 
Traditionally a player's rebounding performance has been measured as the average number of rebounds per game. \citeN{maheswaran-2014} presented a model to quantify the potential to rebound unsuccessful shots in basketball in more detail. Simplified the model considers three phases.  The first phase is the \emph{position} of the players when the shot is made. From the time that the ball is released until it hits the rim, the players will try 
      to move into a better position
-- the \emph{crash} phase. 
      After the crash phase the players have the chance to make the rebound.
The proficiency of a player in rebounding is the measured by the \emph{conversion} -- the third phase.

Both the positioning phase and the crash phase make use of the dominant region (Voronoi diagram) to value the position of the player, i.e., they 
      compute a ``real estate" value of the dominant region of each player both when the shot is made, and when the shot hits the rim.
These values, together with the conversion, are combined into a \emph{rebounding} value.    

\end{enumerate}

\section{Network tools for team performance analysis}
\label{sec:networks}

Understanding the interaction between players is one of the more important and complex problems in sports science. 
Player interaction can give insight into a team's playing style, or be used to assess the importance of individual players to the team. 
Capturing the interactions between individuals is a central goal of social network analysis~\cite{Wasserman1997} and techniques developed in this discipline have been applied to the problem of modelling player interactions.  

An early attempt to use networks for sports analysis was in an entertaining study by \citeN{Gould1979} where they explore all passes made in the 1977 FA Cup Final between Liverpool and Manchester United. 
They studied the simplicial complexes of the passing network and made several interesting observations, including that the Liverpool team had two ``quite disconnected" subsystems and that Kevin Keegan was ``the linchpin of Liverpool". 
However, their analysis, while innovative, 
      did not attract much attention.

In the last decade numerous papers have appeared that apply social network analysis to team sports. Two types of networks have dominated the research literature to date: \emph{passing networks} and \emph{transition networks}. 

\begin{figure} 
\begin{center}
 \includegraphics[width=0.85\textwidth]{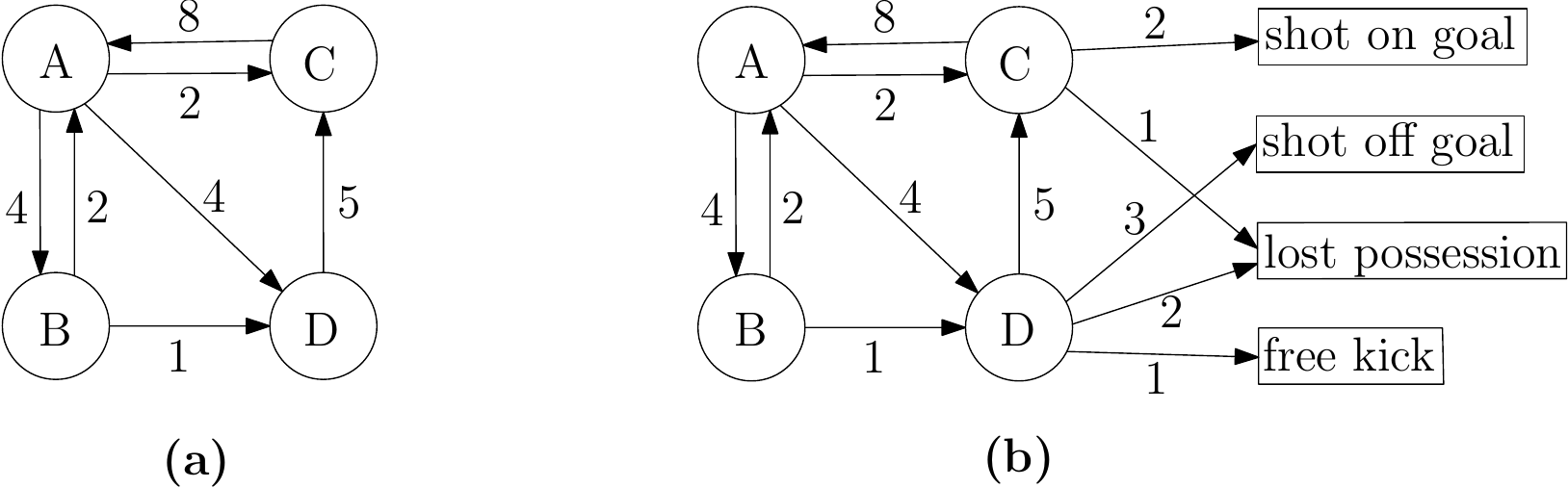}
\end{center}
\caption{(a) A passing network modelling four players $\{A,B,C,D\}$ and the passes between the players. (b) A transition network is a passing network extended with outcomes. For example, twice player $C$ made a shot on goal and once the player lost possession.}
\label{fig:PassNetwork}
\end{figure}

Passing networks have been most frequently studied type in the research field. 
To the best of our knowledge, they were first introduced by \citeN{Passos2011}. 
A passing network is a graph $G=(V,E)$ where each player is modelled as a vertex and two vertices $v_1$ and $v_2$ in $V$ have a directed edge $e=(v_1,v_2)$ from $v_1$ to $v_2$ with integer weight $w(e)$ such that the player represented by vertex $v_1$ has made $w(e)$ successful passes to the player represented by vertex $v_2$. 
A small example of a passing graph is shown in Fig.~\ref{fig:PassNetwork}a. 
Passing networks can be constructed directly from \emph{event logs}, defined in Section~\ref{sec:representation}. A temporal sequence of passes made in a match is encoded as a path within the passing network.
A passing network that is extended with outcomes, as illustrated in Fig.~\ref{fig:PassNetwork}b, is then referred to as a \emph{transition network}.

Many properties of passing networks have been studied, among them \emph{density}, \emph{heterogeneity}, \emph{entropy}, and \emph{Nash equilibria}. However, the most studied measurement is \emph{centrality}.  
We begin by considering centrality and its variants, and then we briefly consider some of the other measures discussed in the literature. 

\subsection{Centrality}
      Centrality measures were introduced in an attempt to determine the key nodes in a network, for example, to identify the most popular persons in a social network or super-spreaders of a disease~\cite{Newman2010}. 
In team sports the aim of using centrality measurements is generally to identify key players, or to estimate the interactivity between team members. For an excellent survey on network centrality see~\citeN{Borgatti2005}.

\subsubsection{Degree centrality}
The simplest centrality measure is the \emph{degree centrality}, which is the number of edges incident to a vertex. 
      For directed networks one usually distinguish between the in-degree and the out-degree centrality.
In sports analysis the out-degree centrality is simply referred to as \emph{centrality} while the in-degree centrality is usually called the \emph{prestige} of a player. 
Some papers do consider both centrality and prestige, see for example~\citeN{Clemente2015c}, but most of the literature has focused on centrality.

\citeN{fewell-2012} considered a transition graph on basketball games where the vertices represented the five traditional player positions (point guard, shooting guard, small forward, power forward, and center), possession origins and possession outcomes. The centrality was computed on the transition graph, split into two outcomes: ``shots" and ``others". The measure was computed on \num{32} basketball games and prior knowledge about the importance of players to the teams involved was compared to the centrality values of the players.   They used degree centrality to compare teams that heavily rely on key players 
      with teams with a more even distribution between their team members.
Unfortunately, the data was not definitive since the overall centrality rankings did not show a strong relation to the teams performance.

\citeN{grund-2012} used degree centrality together with Freeman centralization~\cite{Freeman1978}. 
The idea by Freeman was 
to consider the relative centrality of the most important node in the network. That is, how central is the most central node compared to the centrality of the other nodes in the network. The Freeman centrality is measured as the sum of the differences between the node with the highest degree centrality and all other nodes; divided by a value depending only on the size of the network~\cite{Freeman1978}.
They used an extensive set of \num{283259} passes from \num{760} English Premier League games for their experiments. From a team performance perspective \citeN{grund-2012} set out to answer two hypotheses: (i) increased interaction between players leads to increased team performance; and (ii) increased interaction centralization leads to decreased team performance. 
The latter is strongly connected to centrality and \citeN{grund-2012} went on to show that a high level of centralization decreases team performance.

In a series of recent papers, Clemente~\etal~\shortcite{clemente-2015b,Clemente2014,Clemente2015c,Clemente2015} argue that centrality may 
      recognise how players collaborate, and also the nature and strength of their collaboration.
For example, central midfielders and central defenders usually show higher degree centrality then other players. Some exceptions were shown in~\citeN{Clemente2014a} where the left and right defenders also obtained very high degree centrality. In general goal-keepers and forwards have the lowest centrality measure.

\subsubsection{Betweenness Centrality}
The betweeness centrality of a node is the number of times it lies on the shortest path between two other nodes in the network. Originally it was introduced by~\citeN{Freeman1977} in an attempt to estimate ``a human's potential control of communication in a social network". 

\citeN{Pena2012} claimed that the betweenness centrality measures how the ball-flow between players depends on a particular player and as such provides a measure of the impact of the ``removal" of that player from the game, either by being sent off or by being isolated by the opponents. They also argued that, from a tactical point of view, a team should aim to have a balanced betweenness score for all players.

A centrality measure closely related to the betweenness centrality is flow centrality. The flow centrality is measured by the proportion of the entire flow between two vertices that occur on paths of which a given vertex is a part.

\citeN{Duch2010} considered flow centrality for transition networks where the weight of an edge from a player $v_1$ to a player $v_2$ is equal to the fraction of passes initiated by $v_1$ to reach $v_2$. Similarly, the shooting accuracy for a player (the weight of the edge from the player to the vertex ``shots on goal") is the fraction of shots made by the player that end up on goal. They then studied the flow centrality over all paths that results in a shot. They also take the defensive performance into account by 
      having each player initiate a number of flow paths which is comparable to the number of times the player wins possession of the ball.
The \emph{match performance} of the player is then the normalised value of the logarithm of this combined value. 
They argue that this 
      gives an estimate of the contribution of a single player and also of the whole team. 
The team's match performance value is the mean of the individual player values. Using these values, both for teams and individual players, \citeN{Duch2010} analysed \num{20} games from the football 2008 UEFA European Cup. They claim that their measurements provide ``sensible results that are in agreement with the subjective views of analysts and spectators'', in other words, the better paid players tend to contribute more to the team's performance.

\subsubsection{Closeness Centrality}
      The standard distance metric used in a network is the length (weight or cost) of the shortest path between pairs of nodes. The \emph{closeness centrality} of a node is defined as the inverse of the \emph{farness} of the node, which is the sum of its distance to all other nodes in the network~\citeN{Bavelas1950}. 
\citeN{Pena2012} argued that the closeness score is an estimate of how easy it is to get the ball to a specific player, i.e., a high closeness score indicates a well-connected player within the team. They made a detailed study using the 2010 FIFA World Cup passing data. 
The overall conclusion they reached was that there is a high correlation between high scores in closeness centrality, \emph{PageRank} and clustering (see below), which supports the general perception of the players’ performance reported in the media at the time of the tournament.

\subsubsection{Eigenvector Centrality and \emph{PageRank}}
      The general idea of Eigenvector centrality and PageRank is that the importance of a node depends, not only on its degree, but also on the importance of its neighbours.
\citeN{Cotta2013} used the eigenvector centrality calculated with the power iteration model by \citeN{mises1929}. The measure aims to identify which player has the highest probability to be in possession of the ball after a sequence of passes. They also motivated their measure by a thorough analysis of three games from the 2010 FIFA World Cup, where they argued the correlation between the eigenvector centrality score and the team's performance.

A variant of the eigenvector centrality measure is \emph{PageRank}, which was one of the algorithms used by Google Search to rank web-pages~\cite{Brin1998}. 
The passing graph is represented as an adjacency matrix $A$ where each entry $A_{ji}$ is the number of passes from player $j$ to player $i$. 
In football terms, the \emph{PageRank} centrality index for player $i$ is defined as: $$x_i = p \sum_{j\neq i} \frac{A_{ji}}{L^{out}_j} \cdot x_j + q,$$ where $L^{out}_j=\sum_k A_{jk}$ is the total number of passes made by player $j$, 
$p$ is the parameter representing the probability that a player will decide to give the ball away rather than keep it and shoot, and $q$  is a `free' popularity assigned to each player. 
Note that the \emph{PageRank} score of a player is dependant on the scores of the player's team mates. 
\citeN{Pena2012} argue that the \emph{PageRank} measure gives each player a value that is approximately the likelihood that the player will be in possession of the ball after a fixed number of passes. 
Using data from the 2010 FIFA World Cup, they computed the \emph{PageRank} for the players in the top \num{16} teams, but focused their discussion on the players in the top four teams: Spain, Germany, Uruguay and the Netherlands. 
They showed that the \emph{PageRank} of players in the Dutch and Uruguay teams were more evenly distributed than players from Spain and Germany.
This indicates that no player in those teams has a predominant role in the passing scheme while Xavi Hernandez (Spain) and Bastian Schweinsteiger (Germany) were particularly central to their teams.


\subsection{Clustering Coefficients}
A clustering coefficient is a measure of the degree of which nodes in a network are inclined to cluster together. 
In the sport science literature both the \emph{global} and the \emph{local} clustering coefficients have been applied.
The idea of studying the global cluster coefficient of the players in a team is that it reflects the cooperation between players, that is, the higher coefficient for a player the higher is his cooperation with the other members of the team~\cite{Clemente2014,fewell-2012,Pena2012}. \citeN{fewell-2012} also argued that a high global clustering coefficient indicates that attacking decisions are taken by several players, and thus increases the number of 
      possible attacking paths that have to be assessed by the defence.
\citeN{Pena2012} showed, using the 2010 FIFA World Cup passing data, that Spain, Germany and the Netherlands consistently had very high clustering scores when compared to Uruguay, suggesting that they were extremely well connected teams, in the sense that almost all players contribute. 

\citeN{Cotta2013} considered three games involving Spain from the 2010 FIFA World Cup and used the local clustering coefficient as a player coefficient. They studied how the coefficient changed during the games, and argued for a correlation between the number of passes made by Spain and the local clustering coefficient. They claimed that Spain's clustering coefficient remains high over time, ``indicating the elaborate style of the Spanish team".

It should be noted that it is not completely clear that there is a strong connection between the clustering coefficient and the team performance. For example, \citeN{Pena2012} stated that in their study they did not get any reasonable results and ``will postpone the study of this problem for future work."

\begin{open}
    Various centrality and clustering measures have been proposed to accurately represent some aspect of player or team performance.
    A systematic study reviewing all such measures against predefined criteria, and on a large dataset would be a useful contribution to the field.
\end{open}

\subsection{Density and Heterogeneity}

In general it is believed that stronger collaboration (i.e. more passes) will make the team stronger. This is known as the \emph{density-performance hypothesis}~\cite{balkundi2006}. Therefore a widely-assessed measure of networks is density, which is traditionally calculated as the number of edges divided by the total number of possible edges. This is the density measure used by Clemente et al. in a series of recent papers~\shortcite{Clemente2014b,Clemente2015c,Clemente2015,clemente-2015b}. 
For weighted graphs the measurement becomes slightly more complex. \citeN{grund-2012} defined the \emph{intensity} of a team as the sum of the weighted degrees over all players divided by the total time the team have possession of the ball, i.e., possession-weighted passes per minute.

Related to the density is \emph{passing heterogeneity}, which \citeN{Cintia2015} defined as the standard deviation of the vertex degree for each player in the network.
High heterogeneity of a passing network means that the team tends to coalesce into sub-communities, and that there is a low level of cooperation between players~\cite{Clemente2015}. One interesting observation made by \citeN{Clemente2015} was that the density usually went down in the 2nd half while the heterogeneity went up. 

\begin{open}
    The density-performance hypothesis suggests an interesting metric of team performance.
    Can this hypothesis be tested scientifically?
\end{open}

\subsection{Entropy, Topological Depth, Price-of-Anarchy and Power Law Distributions}

As described above, \citeN{fewell-2012} considered an extended transition graph for basketball games, where they also calculated \emph{player entropy}. Shannon entropy~\cite{shannon-2001} was used to 
     estimate the uncertainty of a ball transition.
The \emph{team entropy} is the aggregated player entropies, which can be computed in many different ways. \citeN{fewell-2012} argue that from the perspective of the opposing team the real uncertainty is the number of options, and computed the team entropy from the transition matrix describing ball movement probabilities across the five standard player positions and the two outcomes.

\citeN{Skinner2010a} showed that passing networks have two interesting properties. 
They identified a correspondence between a basketball transition network and a traffic network, and used insights from the latter to make suppositions about the former. They posited that there may be a difference between the Nash equilibrium of a transition network and the community optimum -- the \emph{Price of Anarchy}. In other words,
      for the best outcome one should not always select the highest-percentage shot. 
A similar observation was made in \citeN{fewell-2012} who noted that the low flow centrality of the most utilised position (point guard) seems to indicate that the contribution of key players can be negatively affected by controlling the ball more often than other players. 
Related to the same concept, \citeN{Skinner2010a} suggested that removing a key player from a match -- and hence the transition network -- may actually \emph{improve} the team performance, a phenomena known as the \emph{Braess' paradox} in network analysis~\cite{braess-2005}.

\section{Data Mining}
\label{sec:data-mining}

The representations and structures described in Sections~\ref{sec:representation}--\ref{sec:networks} are informative in isolation, but may also be the input for more complex algorithmic and probabilistic analysis of team sports. 
In this section, we present a task-oriented survey of the techniques that have been applied, and outline the motivations for these tasks.


\subsection{Applying Labels to Events}

Sports analysts are able to make judgments about events and situations that occur in a match, and apply qualitative or quantitative attributes to that event, for example, to rate the riskiness of an attempted shot on goal, or the quality of a pass.
Event labels such as these can be used to measure player and team performance, and are currently obtained manually by video analysis.
Algorithmic approaches to automatically produce such labels may improve the efficiency of the process.

\citeN{horton-2015} presented a classifier that determines the quality of passes made in football matches by applying a label of \emph{good}, \emph{OK} or \emph{bad} to each pass made, and were able to obtain an accuracy rate of \SI{85.8}{\percent}. 
The classifier uses features that are derived from the spatial state of the match when the pass occurs, including features derived from the dominant region described in Section~\ref{sec:dominant-region}, which were found to be important features to the classifier.

In research by \citeN{beetz-2009}, the approach was to cluster passes, and to then induce a decision tree on each cluster where the passes were labelled as belonging to the cluster or not.
The feature predicates, learned as splitting rules, in the tree could then be combined to provide a description of the important attributes of a given pass.

\citeN{bialkowski-2014a} used the formation descriptors computed with the algorithm presented in \cite{bialkowski-2014b} (see Section~\ref{sub:identifying-formations}) to examine whether formations could accurately predict the identity of a team. 
In the model, a linear discriminant analysis classifier was trained on features describing the team formation, and the learned model was able to obtain an accuracy of \SI{67.23}{\percent} when predicting a team from a league of \num{20} teams.

In \citeN{maheswaran-2012} the authors perform an analysis of various aspects of the rebound in basketball to produce a rebound model.
The rebound is decomposed into three components: the location of the shot attempt; the location where the rebound is taken; and the height of the ball when the rebound is taken. 
Using features derived from this model, a binary classifier was trained to predict whether a missed shot would be successfully rebounded by the offensive team.
The model was evaluated and obtained an accuracy rate of \SI{75}{\percent} in experiments on held-out test data.

\subsection{Predicting Future Event Types and Locations}

The ability to predict how play will unfold given the current game-state has been researched extensively, particularly in the computer vision community.
This has an application in automated camera control, where the camera filming a match must automatically control its pitch, tilt and zoom.
The framing of the scene should ideally contain not just the current action, but the movement of players who can be expected to be involved in future action, and the location of where such future action is likely to occur.

\citeN{kim-2010} considered the problem of modelling the evolution of football play from the trajectories of the players, such that the location of the ball at a point in the near future could be predicted.
Player trajectories were used to compute a dense motion field over the entire playing area, and points of convergence within the motion field identified.
The authors suggest that these points of convergence indicate areas where the ball can be expected to move to with high probability, and the experiments described in the paper demonstrate this with several examples.

\citeN{yue-2014} construct a model to predict whether a basketball player will shoot, pass to one of four teammates, or retain possession. 
The action a player takes is modelled using a multi-class conditional random field. 
The input features to the classifier include latent factors representing player locations which are computed using non-negative matrix factorization, and the experimental results show that these features improve the predictive performance of the classifier.

\citeN{wei-2014} constructed a model to make short-term predictions of which football player will be in possession of the ball after a given interval. They propose a model -- augmented-Hidden Conditional Random Fields (aHCRF) -- that combines local observation features with the hidden layer states to make the final prediction of the player who possess the ball. 
The experimental results show that they are able to design a model that can predict which player will be in possession of the ball after \SI{2}{\second} with \SI{99.25}{\percent} accuracy.

\subsection{Identifying Formations}
\label{sub:identifying-formations}

Sports teams use pre-devised spatial formations as a tactic to achieve a particular objective.
The ability to automatically detect such formations is of interest to sports analysts and coaches.
For example, a coach would be interested in understanding the proportion of time that a team maintains an agreed formation, and also when the team is compelled by the circumstances of the match to alter its formation.
Moreover, when preparing for a future opponent, an understanding of the formation used, and periods where the formation changes would be of interest.

\begin{figure}
    \subfloat[Zone Defence]{
        \includegraphics[width=0.355\linewidth]{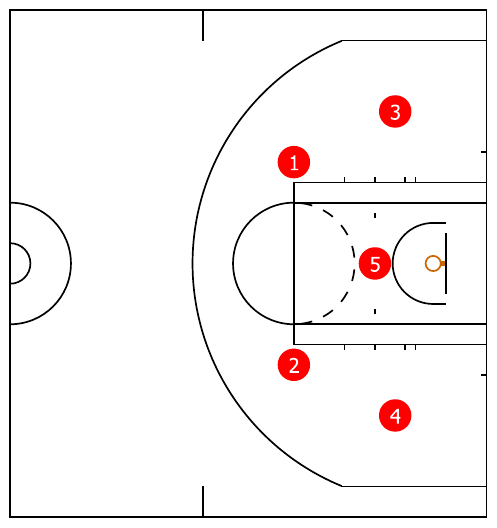}
        \label{fig:bball-zone-defence}}
    \subfloat[Man-marking Defence]{
        \includegraphics[width=0.355\linewidth]{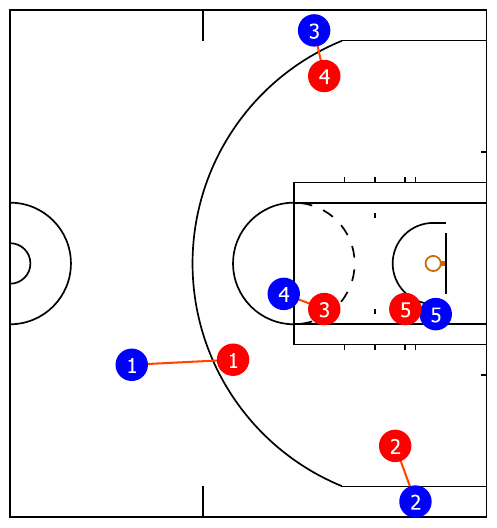}
        \label{fig:bball-man-marking}}
    \subfloat[Back-four Defence]{
        \includegraphics[width=0.285\linewidth]{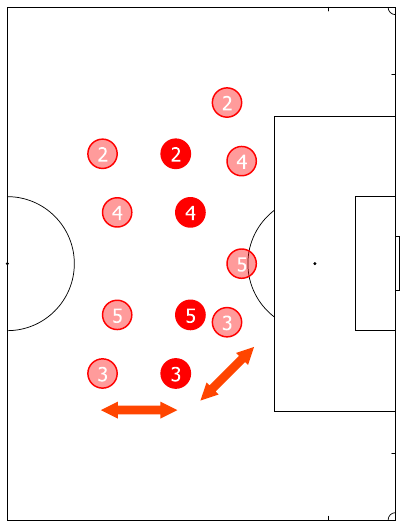}
        \label{fig:football-back-four-defence}}
    \caption{Examples of typical formations used in basketball and football. \protect\subref{fig:bball-zone-defence} The zone defence is spatially anchored to the dimensions of the court and the players positioning is invariant to the phase of play. \protect\subref{fig:bball-man-marking} Defenders who are man-marking will align themselves relative to their opposing player, typically between the attacker and the basket. \protect\subref{fig:football-back-four-defence} The back-four formation in football maintains the alignment of players in the formation, but will move forward and laterally, depending on the phase of play.}
    \label{fig:defense-types}
\end{figure}

A formation is a positioning of players, relative to the location of other objects, such as the pitch boundaries or goal/basket, the players' team-mates, or the opposition players.
Formations may be spatially anchored, for example a zone defence in basketball where players position themselves in a particular location on the playing area, see Fig.~\ref{fig:bball-zone-defence}. 
On the other hand, a formation may vary spatially, but maintain a stable relative orientation between the players in the formation. 
For example, the defensive players in a football team will position themselves in a straight line across the pitch, and this line will move as a group around the pitch, depending on the phase of play, Fig.~\ref{fig:football-back-four-defence}.
Finally, a different type of formation is a \emph{man marking} defence, where defending players will align themselves relative to the attacking players that they are marking, Fig.~\ref{fig:bball-man-marking}.
In this scenario, the locations of defenders may vary considerably, relative to their teammates or to the boundaries of the playing area.

Moreover, the players that fulfil particular roles within a formation may switch, either explicitly through substitutions or dynamically where players may swap roles for tactical reasons.
The following approaches have been used to determine formations from the low-level trajectory signal. 

\citeN{lucey-2013} investigated the assignment of players to roles in field hockey, where teams use a formation of three lines of players arrayed across the field. 
At any time $t$ there is a one-to-one assignment of players to roles, however this assignment may vary from time-step to time-step.
This problem is mathematically equivalent to permuting the player ordering $\vec{p}_t^{\tau}$ using a permutation matrix $\matr{x}_t^{\tau}$ which assigns the players to roles $r_t^{\tau} = \matr{x}_t^{\tau} \vec{p}_t^{\tau}$. 
The optimal permutation matrix $\matr{x}_t^{\tau}$ should minimise the total Euclidean distance between the reference location of each role and the location of the player assigned to the role, and can be computed in closed form using the Hungarian algorithm~\cite{kuhn-1955}.

\citeN{wei-2013} used this approach as a preprocessing step on trajectory data from football matches, and the computed role locations were subsequently used to temporally segment the matches into game phases. 
\citeN{lucey-2014} applied role assignment to basketball players in sequences leading up to three-point shots.
They analysed close to \num{20000} such shots and found that role-swaps involving particular pairs of players in the moments preceding a three-point shot have a significant impact on the probability of the shooter being \emph{open} -- at least \SI{6} feet away from the nearest marker -- when the shot is made.

Furthermore, \citeN{bialkowski-2014b} observed that the role assignment algorithm presented by \citeN{lucey-2013} required a predefined prototype formation to which the players are assigned. 
They consider the problem of simultaneously detecting the reference location of each role in the formation, and assigning players to the formation, using an expectation maximization approach~\cite{dempster-1977}.
The initial role reference locations are determined as the mean position of each player. 
The algorithm then uses the Hungarian algorithm to update the role assignment for each player at each time-step, and then the role reference locations are recomputed according to the role assignment. 
The new locations are used as input for the next iteration, and process is repeated until convergence.

The learned formations for each team and match were then clustered into six formations, and the authors claim that the clustered formations were consistent with expert knowledge of formations used by football teams.
This was validated experimentally by comparing the computed formation with a formation label assigned by an expert, and an accuracy of \SI{75.33}{\percent} was obtained.

In a subsequent paper, \citeN{bialkowski-2014} investigated differences in team strategies when playing home or away, by using formations learned with the role assignment algorithm.
By computing the mean position when teams are playing at home from when they are playing away, they observed that teams defend more deeply when away from home, in other words they set their formation closer to the goal they are defending.

A qualitatively different formation is for players to align themselves with the positions of the opposition players, such as \emph{man-marking} defense used in basketball, see Fig.~\ref{fig:bball-man-marking}.
\citeN{franks-2015} defined a model to determine which defender is marking each attacker. 
For a given offensive player at a given time, the mean location of the defender is modelled as a convex combination of three locations: the position of the attacker, the location of the ball and the location of the hoop. 
The location of a defender, given the observed location of the marked attacker, is modelled as a Gaussian distribution about a mean location. 
The matching between defenders and the attacker that they are marking over a sequence of time-steps is modelled using a Hidden Markov Model, ensuring that the marking assignments are temporally smoothed.

\subsection{Identifying Plays and Tactical Group Movement}

Predefined \emph{plays} are used in many team sports to achieve some specific objective. 
American football uses highly structured plays where the entire team has a role and their movement is highly choreographed.
On the other hand, plays may also be employed in less structured sports such as football and basketball when the opportunity arises, such as the \emph{pick and roll} in basketball or the \emph{one-two} or \emph{wall pass} in football.
Furthermore, teammates who are familiar with each other's playing style may develop ad-hoc productive interactions that are used repeatedly, a simple example of which is a sequence of passes between a small group of players.
Identification of plays is a time-consuming task that is typically carried out by a video analyst, and thus a system to perform the task automatically would be useful.

An early attempt in this direction attempted to recognise predefined plays in American football~\cite{intille-1999}. 
They model a play as a choreographed sequence of movements by attacking players, each trying to achieve a local goal, and in combination achieve a group goal for the play.
The approach taken was to encode predefined tactical plays using a temporal structure description language that described a local goal in terms of a sequence of actions carried out by an individual player. 
These local goals were identified in the input trajectories using a Bayesian belief network. 
A second belief network then identified whether a global goal had been achieved based on the detected local goals -- signifying that the play has occurred. 

Two papers by Li~\etal investigated the problem of identifying group motion, in particular the type of offensive plays in American football. \citeN{li-2009} presented the Discriminative Temporal Interaction Network (DTIM) framework to characterise group motion patterns. The DTM is a temporal interaction matrix that quantifies the interaction between objects at two given points in time.
For each predefined group motion pattern -- a play -- a multi-modal density was learned using a properly defined Riemannian metric, and a MAP classifier was then used to identify the most likely play for a given input set of trajectories. 
The experiments demonstrated that the model was able to accurately classify sets of trajectories into five predefined plays, and outperformed several other common classifiers for the task.
This model has the advantage of not requiring an \emph{a priori} definition of each player's movement in the play, as required in \citeN{intille-1999}.

\citeN{li-2010} considered group motion segmentation, where a set of unlabelled input trajectories are segmented into the subset that participated in the group motion, and those that did not.
The problem was motivated by the example of segmenting a set of trajectories into the set belonging to the offensive team (who participated in the play) and the defensive team (who did not).
The group motion is modelled as a dynamic process driven by a \emph{spatio-temporal driving force} -- a densely distributed motion field over the playing area. The driving force is modelled as a $3 \times 3$ matrix $F(t_0, t_f, x, y)$ such that $X(t_f) = F(t_0, t_f, x, y)X(t_0)$. 
Thus, an object located at $X(t_0)$ at time $t_0$ will be driven to $X(t_f)$ at time $t_f$.
Using Lie group theory \cite{rossmann-2002}, a Lie algebraic representation $f$ of $F$ is determined with the property that the space of all $f$s is linear, and thus tractable statistical tools can be developed upon $f$. 
A Gaussian mixture model was used to learn a fixed number of driving forces at each time-step, which was then used to segment the trajectories.

There has been number of diverse efforts to identify commonly occurring sequences of passes in football matches.
In \citeN{borrie-2002}, the pitch is subdivided into zones and sequences of passes are identified by the zones that they start and terminate in. 
A possession can thus be represented by a string of codes representing each pass by source and target zone, and with an elapsed time between them. 
They introduce \emph{T-pattern} analysis which is used to compute possessions where the same sequence of passes are made with consistent time intervals between each pass, and frequently occurring patterns could thus be identified.
\citeN{camerino-2012} also used T-pattern analysis on pass strings, however the location of passes was computed relative to the formation of the team in possession, e.g. 
between the defense and midfield, or in front of the attacking line.

An algorithm to detect frequently occurring sequences of passes was presented in \citeN{gudmundsson-2014}. 
A suffix tree~\cite{weiner-1973} was used as a data structure $D$ to store sequences of passes between individual players. 
A query $(\tau, o)$ can then be made against $D$ that returns all permutations of $\tau$ players such that the ball is passed from a player $p_1$ to $p_{\tau}$, via players $p_2, \ldots, p_{\tau - 1}$ at least $o$ times, and thus determine commonly used passing combinations between players.

\citeN{vanhaaren-2015} 
     considered the problem of finding patterns in offensive football strategies.
The approach taken was to use inductive logic programming to learn a series of clauses describing the pass interactions between players during a possession sequence that concludes with a shot on goal. 
The passes were characterised by their location on the pitch, and a hierarchical model was defined to aggregate zones of the pitch into larger regions. 
The result is a set of rules, expressed in first-order predicate logic, describing the frequently-occurring interaction sequences.

Research by \citeN{wang-2015} also aimed to detect frequent sequences of passing. 
They claim that the task of identifying tactics from pass sequences is analogous to identifying topics from a document corpus, and present the Team Tactic Topic Model (T$^3$M) based on Latent Dirichlet Allocation \cite{blei-2003}.
Passes are represented as a tuple containing an order-pair of the passer and receiver, and a pair of coordinates representing the location where the pass was received.
The T$^3$M is an unsupervised approach for learning common tactics, and the learned tactics are coherent with respect to the location where they occur, and the players involved.

\subsection{Temporally Segmenting the Game}

Segmenting a match into phases based on a particular set of criteria is a common task in sports analysis, as it facilitates the retrieval of important phases for further analysis.
The following paragraphs describe approaches that have been applied this problem for various types of criteria.

Hervieu~\etal~\shortcite{hervieu-2009,hervieu-2011} present a framework for labelling phases within a handball match from a set of predefined labels representing common attacking and defensive tactics.
The model is based on a hierarchical parallel semi-Markov model (HPaSMM) and is intended to model the temporal causalities implicit in player trajectories. 
In other words, modelling the fact that one player's movement may cause another player to subsequently alter their movement. 
The upper level of the hierarchical model is a semi-Markov model with a state for each of the defined phase labels, and within each state the lower level is a parallel hidden Markov model for each trajectory. The duration of time spent in each upper level state is modelled using a Gaussian mixture model.
In the experiments, the model was applied to a small dataset of handball match trajectories from the $2006$ Olympics Games final, and resulted in accuracy of \SI{92}{\percent} accuracy on each time-step, compared to the ground truth provided by an expert analyst. 
The model exactly predicted the sequence of states, and the misclassifications were all the result of time-lags when transitioning from one state to the subsequent state.


\citeN{Perse2009a} investigated segmentation of basketball matches.
A framework with two components was used, the first segmented the match duration into sequences of offensive, defensive or time-out phases.
The second component identified basic activities in the sequence by matching to a library of predefined activities, and the sequences of activities were then matched with predefined templates that encoded known basketball plays.


\citeN{wei-2013} considered the problem of automatically segmenting football matches into distinct game phases that were classified according to a two-level hierarchy, using a decision forest classifier. 
At the top level, phases were classed as being \emph{in-play} or a \emph{stoppage}. \emph{In-play} phases were separated into highlights or non-highlights; and \emph{stoppages} were classified by the reason for the stoppage: \emph{out for corner}, \emph{out for throw-in}, \emph{foul} or \emph{goal}. 
The classified sequences were subsequently clustered to find a team's most probable method of scoring and of conceding goals.


In a pair of papers by Bourbousson~\etal~\shortcite{bourbousson-2010a,bourbousson-2010b}, the spatial dynamics in basketball was examined using relative-phase analysis. In \citeN{bourbousson-2010a}, the spatial relation between dyads of an attacking player and their marker were analysed.
In \citeN{bourbousson-2010b}, the pairwise relation between the centroid of each team was used, along with a \emph{stretch index} that measured the aggregate distance betweens players and their team's centroid. 
A Hilbert transformation was used to compute the relative phase in the $x$ and $y$ direction of the pairs of metrics.
Experimental results showed a strong in-phase relation between the various pairs of metrics in the matches analysed, suggesting individual players and also teams move synchronously.
The authors suggest that the spatial relations between the pairs are consistent with their prior knowledge of basketball tactics.

\citeN{frencken-2011} performed a similar analysis of four-a-side football matches. They used the centroid and the convex hull induced by the positions of the players in a team to compute metrics, for example the distance in the $x$ and $y$ direction of the centroid, and the surface area of the convex hull.
The synchronized measurements for the two teams were modelled as coupled oscillators, using the HKB-model~\cite{haken-1985}. 
Their hypothesis was that the measurements would exhibit in-phase and anti-phase coupling sequences, and that the anti-phase sequences would denote game-phases of interest.
In particular, the authors claim that there is a strong linear relationship between the $x$-direction of the centroid of the two teams, and that phases where the centroid's $x$-directions cross are indicative of unstable situations that are conducive to scoring opportunities. 
They note that such a crossing occurs in the build up to goals in about half the examples.

\begin{open}
    Coaches and analysts are often interested in how the \emph{intensity} of a match varies over time, as periods of high intensity tend to be present more opportunities and threats.
    It is an interesting open problem to determine if it is possible to compute a measure of intensity from spatio-temporal data, and thus be able to determine high-intensity periods.
\end{open}

\section{Performance Metrics}
\label{sec:performance-metrics}

Determining the contribution of the offensive and defensive components of team play has been extensively researched, particularly in the case of basketball which has several useful properties in this regard.
For example, a basketball match can be easily segmented into a sequence of \emph{possessions} -- teams average around \num{92} possessions per game~\cite{kubatko-2007} -- most of which end in a shot at goal, which may or may not be successful.
This segmentation naturally supports a variety of offensive and defensive metrics \cite{kubatko-2007}, however the metrics are not spatially informed, and intuitively, spatial factors are significant when quantifying both offensive and defensive performance.
In this section we survey a number of research papers that use spatio-temporal data from basketball matches to produce enhanced performance metrics.

\subsection{Offensive Performance}

Shooting effectiveness is the likelihood that a shot made will be successful, and \emph{effective field goal percentage} (EFG) is a de-facto metric for offensive play in basketball~\cite{kubatko-2007}. 
However, as \citeN{chang-2014} observe, this metric confounds the efficiency of the shooter with the difficulty of the shot. 
Intuitively, spatial factors such as the location where a shot was attempted from, and the proximity of defenders to the shooter would have an impact of the difficulty of the shot.
This insight has been the basis of several efforts to produce metrics that provide a more nuanced picture of a player or team's shooting efficiency.

Early work in this area by \citeN{reich-2006} used shot chart data (a list of shots attempted, detailing the location, time, shooter and outcome of each shot). 
The paper contained an in-depth analysis of the shooting performance of a single player -- Sam Cassell of the Minnesota Timberwolves -- over the entire $2003\mbox{/}2004$ season. 
A vector of boolean-valued predictor variables was computed for each shot, and linear models fitted for shot frequency, shot location and shot efficiency. 
By fitting models on subsets of the predictor variables, the authors analysed the factors that were important in predicting shot frequency, location and efficiency.

\citeN{miller-2014} investigated shooting efficiency by using vectors computed with non-negative matrix factorization to represent spatially distinct shot-types, see Section~\ref{sub:matrix-factorization}. 
The shooting factors were used to estimate spatial shooting efficiency surfaces for individual players. 
The efficiency surfaces could then be used to compute the probability of a player making a shot conditioned on the location of the shot attempt, resulting in a spatially-varying shooting efficiency model for each individual player.

\citeN{cervone-2014a} present \emph{expected possession value} (EPV), a continuous measure of the expected points that will result from the current possession. EPV is thus analogous to a ``stock ticker'' that provides a valuation of the possession at any point in time during the possession. 
The overall framework consists of a macro-transition model that deals with game-state events such as passes, shots and turnovers, and micro-transition model that describes player movement within a phase when a single player is in possession of the ball. Probability distributions, conditioned on the spatial layout of all players and the ball, are learned for the micro- and macro-transition models. 
The spatial effects are modelled using non-negative matrix factorization to provide a compact representation that the authors claim has the attributes of being computationally tractable with good predictive performance.
The micro- and macro-transition models are combined in a Markov chain, and from this the expected value of the end-state -- scoring $0$, $2$ or $3$ points -- can be determined at any time during the possession.
Experimental results in the paper show how the EPV metric can support a number of analyses, such as \emph{EPV-Added} which compares an individual player's offensive value with that of a league-average player receiving the ball in the same situation; or \emph{Shot Satisfaction} which quantifies the satisfaction (or regret) of a player's decision to shoot, rather than taking an alternative option such as passing to a teammate.

\citeN{chang-2014} introduces another spatially-informed measure of shooting quality in basketball: \emph{Effective Shot Quality} (ESQ). 
This metric measures the value of a shot, were it to be taken by the league-average player. 
ESQ is computed using a learned least-squares regression function whose input includes spatial factors such as the location of the shot attempt, and the proximity of defenders to the shooter. 
Furthermore, the authors introduce EFG+, which is calculated by subtracting ESQ from EFG. 
EFG+ is thus an estimate of how well a player shoots relative to expectation, given the spatial conditions under which the shot was taken.

A further metric, \emph{Spatial Shooting Effectiveness}, was presented by \citeN{shortridge-2014}. 
Using a subdivision of the court, an empirical Bayesian scoring rate estimator was fitted using the neighbourhood of regions to the shot location. 
The spatial shooting effectiveness was computed for each player in each region of the subdivision, and is the difference between the points-per-shot achieved by the player in the region and the expected points-per-shot from the estimator.
In other words, it is the difference between a player's expected and actual shooting efficiency, and thus measures how effective a player is at shooting, relative to the league-average player.


\citeN{lucey-2014a} considered shooting efficiency in football. 
They make a similar observation that the location where a shot is taken significantly impacts the likelihood of successfully scoring a goal.
The proposed model uses logistic regression to estimate the probability of a shot succeeding -- the \emph{Expected Goal Value} (EGV). 
The input features are based on the proximity of defenders to the shooter and to the path the ball would take to reach the goal; the location of the shooter relative to the lines of players in the defending team's formation; and the location where the shot was taken from.
The model is empirically analysed in several ways.
The number of attempted and successful shots for an entire season is computed for each team in a professional league, and compared to the expected number of goals that the model predicts, given the chances. 
The results are generally consistent, and the authors are able to explain away the main outliers. 
Furthermore, matches where the winning team has fewer shots at goal are considered individually, and the expected goals under the model are computed. 
This is shown to be a better predictor of the actual outcome, suggesting that the winning team was able to produce fewer -- but better -- quality chances. 

\subsection{Defensive Performance}

Measures of defensive performance have traditionally been based on summary statistics of \emph{interventions} such as blocks and rebounds in basketball~\cite{kubatko-2007} and tackles and clearances in football. 
However, \citeN{goldsberry-2013} observed that, in basketball the defensive effectiveness ought to consider factors such as the spatial dominance by the defence of areas with high rates of shooting success; the ability of the defence to prevent a shot from even being attempted; and secondary effects in the case of an unsuccessful shot, such as being able to win possession or being well-positioned to defend the subsequent phase.

In order to provide a finer-grained insight into defensive performance, \citeN{goldsberry-2013} presented \emph{spatial splits} that decompose shooting frequency and efficiency into a triple consisting of close-range, mid-range and 3-point-range values. 
The offensive half-court was subdivided into three regions, and the shot frequency and efficiency were computed separately for shots originating in each region.
These offensive metrics were then used to produce defensive metrics for the opposing team by comparing the relative changes in the splits for shots that an individual player was defending to the splits for the league-average defender.

An alternate approach to assessing the impact of defenders on shooting frequency and efficiency was taken by Franks~\etal~\shortcite{franks-2015,franks-2015a}. 
They proposed a model that quantifies the effectiveness of man-to-man defense in different regions of the court. 
The proposed framework includes a model that determines {who's marking whom} by assigning each defender to an attacker. 
For each attacker, the canonical position for the defender is computed, based on the relative spatial location of the attacker, the ball and the basket. 
A hidden Markov model is used to compute the likelihood of an assignment of defenders to attackers over the course of a possession, trained using the expectation maximization algorithm~\cite{dempster-1977}.
A second component of the model learned spatially coherent shooting type bases using non-negative matrix factorization on a shooting intensity surface fitted using a log-Gaussian Cox process. 
By combining the assignment of markers and the shot type bases, the authors were able to investigate the extent to which defenders inhibit (or encourage) shot attempts in different regions of the court, and the degree to which the efficiency of the shooter is affected by the identity of the marker.

Another aspect of defensive performance concerns the actions when a shot is unsuccessful, and both the defence and attack will attempt to \emph{rebound} the shot to gain possession.
This was investigated by \citeN{maheswaran-2014} where they deconstructed the rebound into three components: \emph{positioning}; \emph{hustle} and \emph{conversion}, described in Section~\ref{sssec:DR}.
Linear regression was used to compute metrics for player's \emph{hustle} and \emph{conversion}, and experimental results showed that the top-ranked players on these metrics were consistent with expert consensus of top-performing players.

On the other hand, \citeN{wiens-2013} performs a statistical evaluation of the options that players in the offensive team have when a shot is made in basketball. 
Players near the basket can either \emph{crash the boards} -- move closer to the basket in anticipation of making a rebound -- or retreat in order to maximise the time to position themselves defensively for the opposition's subsequent attack. 
The model used as factors the players' distance to the basket, and proximity of defenders to each attacking player. 
The experimental results suggested that teams tended to retreat more than they should, and thus a more aggressive strategy could improve a team's chances of success.

The analysis of defense in football would appear to be a qualitatively different proposition, in particular because scoring chances are much less frequent. To our knowledge, similar types of analysis to those presented above in relation to basketball have not been attempted for football.

\begin{open}
    There has been significant research into producing spatially-informed metrics for player and team performance in basketball, however there has been little research in other sports, particularly football.
    It is an open research question whether similar spatially-informed sports metrics could be developed for football.
\end{open}

\section{Visualisation}
\label{sec:visualisation}


To communicate the information extracted from the spatio-temporal data, visualization tools are required. 
      For real-time data the most common approach is so-called \emph{live covers}. This is usually a website that comprise of a text panel that lists high-level updates of the key events in the game in almost real time, and several graphics showing basic information about the teams and the game.
Live covers are provided by leagues (e.g. NHL, NBA and Bundesliga), media (e.g. ESPN) and even football clubs (e.g. Liverpool and Paris Saint-Germain). For visualizing aggregated information the most common approach is to use heat maps. Heat maps are simple to generate, are intuitive, and can be used to visualize various types of data. Typical examples in the literature are, visualizing the spread and range of a shooter (basketball) in an attempt to discover the best shooters in the NHL~\cite{goldsberry-2012} and visualizing the shot distance (ice hockey) using radial heat maps~\cite{Pileggi2012a}. 
Two recent attempts to provide more extensive visual analytics systems have been made by~\citeN{perin-2013} and Janetzko~\etal~\shortcite{Stein2015} .

\citeN{perin-2013} developed a system for visual exploration of phases in football.  The main interface is a timeline and \emph{small multiples} providing an overview of the game. A \emph{small multiple} is a group of similar graphs or charts designed to simplify comparisons between them.  The interface also allows the user to select and further examine the \emph{phases} of the game. A phase is a sequence of actions by one team bounded by the actions in which they first win, and then finally lose possession.
A selected phase can be displayed and the information regarding a phase is aggregated into a sequence of views, where each view only focus on a specific action (e.g. a long ball or a corner). The views are then connected to show a whole phase, using various visualization tools such as a passing network, a time line and sidebars for various detailed information.


In two papers Janetzko~\etal~\shortcite{janetzko-2014,Stein2015} present a visual analysis system for interactive recoginition of football patterns and situations. Their system tightly couple the visualization system with data mining techniques. The system includes a range of visualization tools (e.g., parallel coordinates and scalable bar charts) to show the ranking of features over time and plots the change of game play situations, attempting to support the analyst to interpret complex game situations. Using classifiers they automatically detected the most common situations and introduced semantically-meaningful features for these. The exploration system also allows the user to specify features for a specific situations and then perform a similarity search for similar situations. 
\begin{open} \label{probel:visualization}
  The area of visual interfaces to support team sports analytics is a developing area of research. Two crucial gaps are large user studies with the aim to (1) explore the analytical questions that experts need support for, and (2) which types of visual analytical tools can be understood by experts? 
\end{open}


\section{Conclusion}
\label{sec:conclusion}

The proliferation of optical and device tracking systems in the stadia of teams in professional leagues in recent years have produced a large volume of player and ball trajectory data, and this has subsequently enabled a proliferation of research efforts across a variety of research communities.
To date, a diversity of techniques have been brought to bear on a number of problems, however there is little consensus on the key research questions or the techniques to use to address them.
Thus, we believe that this survey of the current research questions and techniques is a timely contribution to the field.

This paper surveys the recent research into team sports analysis that is primarily based on spatio-temporal data, and describes a framework in which the various research problems can be categorised.
We believe that the structured approach used in this survey reflects a useful classification for the research questions in this area.
Moreover, the survey should be useful as a concise introduction for researchers who are new to the field.

\bibliographystyle{plainnat}
\bibliography{References-Survey}

\end{document}